\documentclass[10pt,a4paper]{article}
\usepackage{graphpap,epsfig,amssymb,graphicx,textcomp}
\begin{document}

\begin{center}
{\Large {\bf Driven spin wave modes in XY ferromagnet:
Nonequilibrium phase transition}}\end{center}

\vskip 1cm

\begin{center}{\it Muktish Acharyya}\\
{\it Department of Physics, Presidency University,}\\
{\it 86/1 College Street, Calcutta-700073, INDIA}\\
{E-mail:muktish.physics@presiuniv.ac.in}\end{center}

\vskip 1cm

\noindent {\bf Abstract:} The dynamical responses of 
XY ferromagnet driven by linearly polarised
propagating and standing magnetic field wave have been studied by Monte Carlo
simulation in three dimensions. In the case of propagating magnetic field 
wave (with specified amplitude, frequency and the wavelength), the low 
temperature dynamical mode is a propagating spin wave and the system becomes
structureless (or random) in the high temperature. A dynamical symmetry breaking
phase transition is observed at a finite (nonzero) temperature. This symmetry
breaking is confirmed by studying the statistical distribution of the angle
of the spin vector. The dynamic nonequilibrium transition temperature was found
to decrease as the amplitude of the propagating magnetic field wave increased.
A comprehensive phase boundary is drawn in the plane formed by temperature
and amplitude of propagating field wave. The phase boundary was observed to
shrink (in the low temperature side) for longer wavelength of the propagating
magnetic wave. In the case of standing magnetic field wave, the low temperature
excitation is a standing spin wave which becomes structureless (or random) in
the high temperature. Here also, like the case of propagating magnetic wave,
a dynamical symmetry breaking nonequilibrium phase transition was observed.
A comprehensive phase boundary was drawn. Unlike the case of propagating 
magnetic wave, the phase boundary does not show any systematic variation with
the wavelength of the standing magnetic field wave. In the limit of 
vanishingly small
amplitude of the field, the phase boundaries approach the recent Monte Carlo
estimate of equilibrium transition temperature.

\vskip 2cm


\newpage

\noindent {\bf I. Introduction:}

The nonequilibrium responses of Ising ferromagnet to an oscillating
(in time but uniform over the space) 
magnetic field is an interesting
field of modern research\cite{rmp,marev}. The 
nonequilibrium phase transition is one major focus of the investigation.
Some important studies may be reported below.
The existence of the growth of correlation near the transition was reported
\cite{sides}. 
The bulk and surface critical behaviours were studied recently\cite{park}
and those are found to belong to different universality class. The anomalous metamagnetic
fluctuations near the transition were studied recently\cite{reigo}. This
study was supported by Monte Carlo simulation\cite{rik}. 
Experimentally, a notable transient behaviour was found\cite{berger}, in the 
uniaxial cobalt film,  
for the time period of the field which is faster than a critical 
time. This is related to the
the existence of first order transition.
All these studies, mentioned above, are signatures of the current interest
in the field of nonequilibrium responses of ferromagnets driven by time
varying external magnetic field.

One common and important feature of 
the above mentioned studies, is the time dependence
of the external magnetic field, which keeps the system 
far away from the equilibrium.
However, recently the interests have been taken 
in the case, where the driving magnetic
field has both spatial and temporal variations. This spatio-temporal variations
have been incorporated as the propagating and standing magnetic field wave
with specified amplitude, frequency and wavelength. The Ising ferromagnet
driven by propagating magnetic field wave has been studied\cite{prop}
by Monte Carlo simulation. The low temperature pinned (or frozen)
phase was observed, where almost all the spins are parallel and 
remain in a frozen state. 
Above a certain critical temperature, the coherent propagation of spin bands
is observed. The nonequilibrium dynamic transition is found and comprehensive
phase boundary was obtained. In the case of standing magnetic wave
\cite{ajay}, standing
spin band modes (in the high temperature) are observed. The similar studies
are performed with standing magnetic wave
\cite{rfim} in random field Ising ferromagnet and the
 exact mathematical form of the phase
boundary is found, the breathing and spreading
transitions are found in Ising ferromagnet driven 
by spherical magnetic wave \cite{ma14}, 
 the dynamic transitions are studied in Blume-Capel model \cite{blume} 
driven by propagating and standing magnetic wave. The 
nonequilibrium multiple phase
transition was observed\cite{meta} in Ising metamagnet driven by propagating
magnetic field wave. All these studies mentioned above are perfomed in the
discrete (Ising, Blume-Capel etc.) spin models.

The continuous ferromagnetic spin model like XY model has a very rich variety
of behaviours. The exsistence of a very special kind of phase  
without any long-range order was first proposed by Kosterlitz and Thouless
\cite{kosterlitz} in planar magnets like two dimensional XY ferromagnets.
After this remarkable discovery, the XY ferromagnet has drawn much attention
of the researchers. The critical dynamics of two dimensional XY ferromagnet
was studied\cite{landau} by Monte Carlo simulation. 
The surface critical behaviour was studied\cite{binder} in the XY model by
Monte Carlo simulation.
The Monte Carlo simulation
was perfomed\cite{rastelli} in planar rotator model with symmetry breaking
field. The vortex glass transition was found\cite{olsson} is three dimensional
XY model by Monte Carlo simulation. Recently, the quantum phase transition
was found\cite{guimaraes} in quantum XY model. The results 
of the
critical properties of frustrated
quasi two dimensional XY like antiferromagnet were reported\cite{lapa}.
The magnetic properties of classical XY spin dimer in planar magnetic field
were studied recently\cite{ciftza}. All the studies mentioned in this paragraph
deal mainly with the {\it equilibrium} properties of XY model.

The {\it nonequilibrium} critical dynamics in XY model was also studied
\cite{berthier}. Nonequilibrium quantum phase transition using $C^*$ algebra
was studied recently \cite{ajisaka}. Nonequilibrium phase transition in 
XY model with long range interaction was also studied \cite{teles}.
The dynamical phase transition in anisotropic XY ferromagnet driven by
oscillating (in time but uniform over space) magnetic field was studied
\cite{yasui}. However, as far as the knowledge of this author is concerned,
the nonequilibrium responses of XY ferromagnet to a magnetic field having
the spatio-temporal variation, has not been studied so far.

In this paper, the nonequilibrium phase transition in XY ferromagnet,
driven by propagating and standing magnetic field wave is
studied by Monte Carlo simulation in three dimensions. The paper is organised
as follows: Section-II describes the model and the Monte Carlo simulation
method, the numerical results (with diagrams) are reported in section-III and
the paper ends with a summary in section-IV.

\vskip 1cm

\noindent {\bf II. Model and simulation}

The time dependent Hamiltonian of XY ferromagnet driven by a field having the spatio-temporal
variation is expressed as

\begin{equation}
H(t) = -J\sum \sum {\rm cos}(\theta(x,y,z,t)-\theta(x',y',z',t)) - \sum h(x,y,z,t)
{\rm cos}(\theta(x,y,z))
\end{equation}

First term represents the distinct sum of all interactions between spins at site x,y,z with
its neighbouring (nearest) site x', y', z' at any instant $t$. The x and y components of the spin
vector are represented by $s_x$ and $s_y$ respectively and $s=\sqrt{s_x^2+
s_y^2}=1$ here.
The Spatio-temporal variations of the driving magnetic field have both
(i) propagating wave form $h(x,y,z,t)=H {\rm cos}[2\pi(ft-z/{\lambda})]$
and (ii) standing wave form $h(x,y,z,t)= H {\rm sin}(2\pi f t) {\rm sin}
(2\pi z/{\lambda})$. The magnetic field wave propagates (or extends) along
the z-direction and the field oscillates along the x-direction. 
The magnetic field wave is linearly (along x direction) polarised.
$J$ is the
ferromagnetic ($J > 0$) interaction strength. The magnitude of the field
($h(x,y,z,t)$) is measured in the unit of $J$. A cubic lattice of size $L$
(=20 here)
is considered. It may be noted here that the space dimensions of the 
system (lattice) are three
(cubic) and the dimensions of the spin vector are two (XY model). The boundary 
conditions are taken as periodic in all three directions of the lattice.

The simulation starts from a random initial spin configuration corresponding to
a very high temperature phase.
At any finite temperature $T$ (measured in the unit of $J/k$, where $k$ is Boltzmann
constant), a site (say x,y,z) is chosen randomly 
(at any instant $t$) having an initial spin configuration (represented
by an angle $\theta_i(x,y,z,t)$). A new configuration of the spin 
(at site x,y,z and at the same instant $t$) is also chosen
(represented by $\theta_f(x,y,z,t)$) randomly. The change in energy
($\delta H(t)$) due to the change in configuration (angle)
of spin (from $\theta_i(x,y,z,t)$ to $\theta_f(x,y,z,t)$) is calculated 
from equation (1). The probability of accepting
the new configuration is calculated from the Metropolis formula\cite{metro}

\begin{equation}
P_f = {\rm Min}[{\rm exp}({{-\delta H(t)} \over {kT}}), 1].
\end{equation}

An uniformly distributed random number ($r=[0,1]$) is chosen. 
The chosen site is assigned to the new
spin configuration $\theta_f(x,y,z,t')$ (for the next instant $t'$)
if $r \leq P_f$. In this way, $L^3$ number of sites are updated randomly. $L^3$ number of such
random updates defines a unit time step and is called Monte Carlo step per site (MCSS). The time in this simulation is measured in the unit of MCSS.
Throughout the study the system size $L(=20)$ and frequency $f(=0.01)$ of the magnetic field wave
are kept fixed. The total length of simulation is $1.5\times10^5$ MCSS, out of which initial
$0.5\times10^5$ MCSS times are discarded. All statistical quantities are calculated over rest
$10^5$ MCSS. 

The instantaneous components of magnetisations are 
$Mx(t)= {1 \over {L^3}} \sum s_x (x,y,z,t)\\ 
= {1 \over {L^3}} \sum {\rm cos} (\theta (x,y,z,t))$ 
and 
$My(t)= {1 \over {L^3}} \sum s_y(x,y,z,t)
= {1 \over {L^3}} \sum {\rm sin}(\theta(x,y,z,t)$.
The components of the time averaged magnetisation
over a full cycle (in time) of the magnetic field wave are defined as,
$Qx = f \oint Mx(t) dt$ and
$Qy = f \oint My(t) dt$. 
Since the frequency $f$ is taken equal to 0.01, 100 MCSS are 
required to have a complete temporal cycle
of the magnetic field wave. In $10^5$ MCSS, 1000 such cycles are present. The $Qx$ and $Qy$ are
calculated as the average over 1000 cycles. 
The variances of $Qx$ and $Qy$ are defined as 
$Var(Qx) = L^3(<Qx^2> - <Qx>^2)$ and
$Var(Qy) = L^3(<Qy^2> - <Qy>^2)$. The time averaged energy over the full cycles of the magnetic field
wave is $E = f \oint H(t) dt$ and the dynamic specific heat is defined as $C= {{dE} \over {dT}}$.

\vskip 1.0 cm

\noindent {\bf III. Results:}

\vskip 0.5cm

\noindent {\it (a) Propagating wave:}

\begin{figure}[h]
\begin{center}
\begin{tabular}{c}
\resizebox{7cm}{!}{\includegraphics[angle=0]{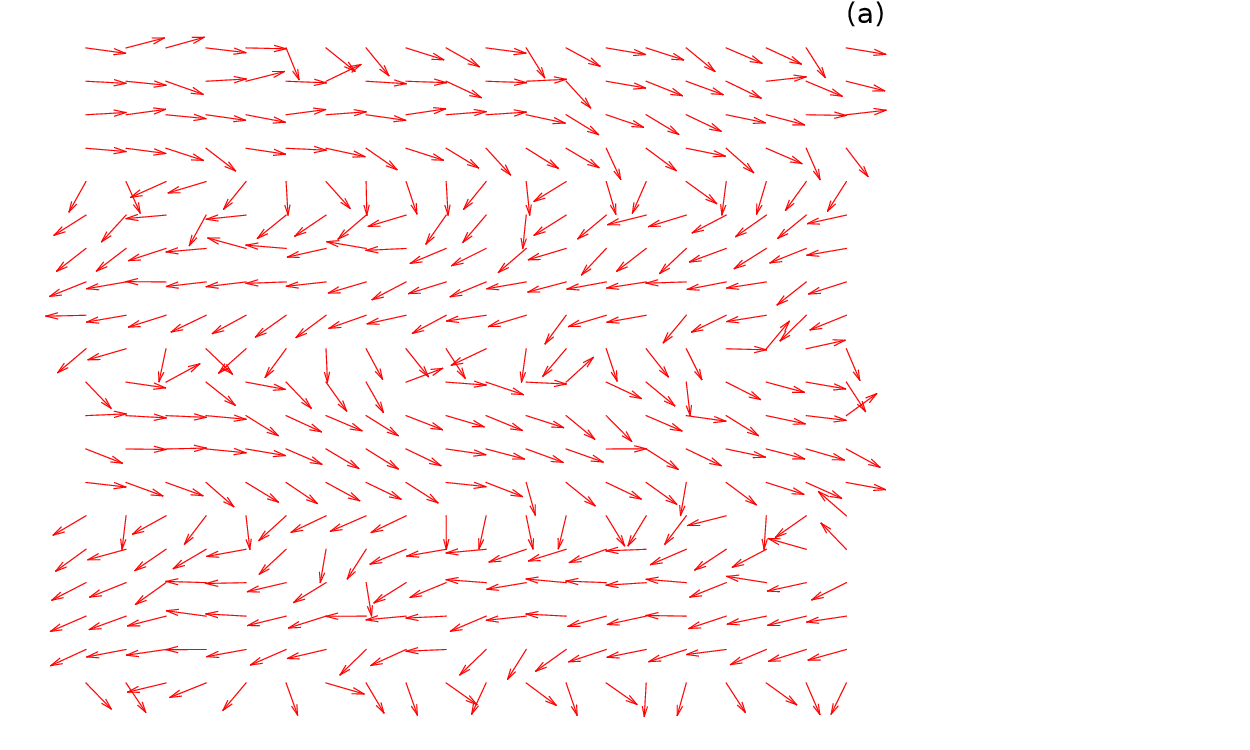}}
\\
\resizebox{7cm}{!}{\includegraphics[angle=0]{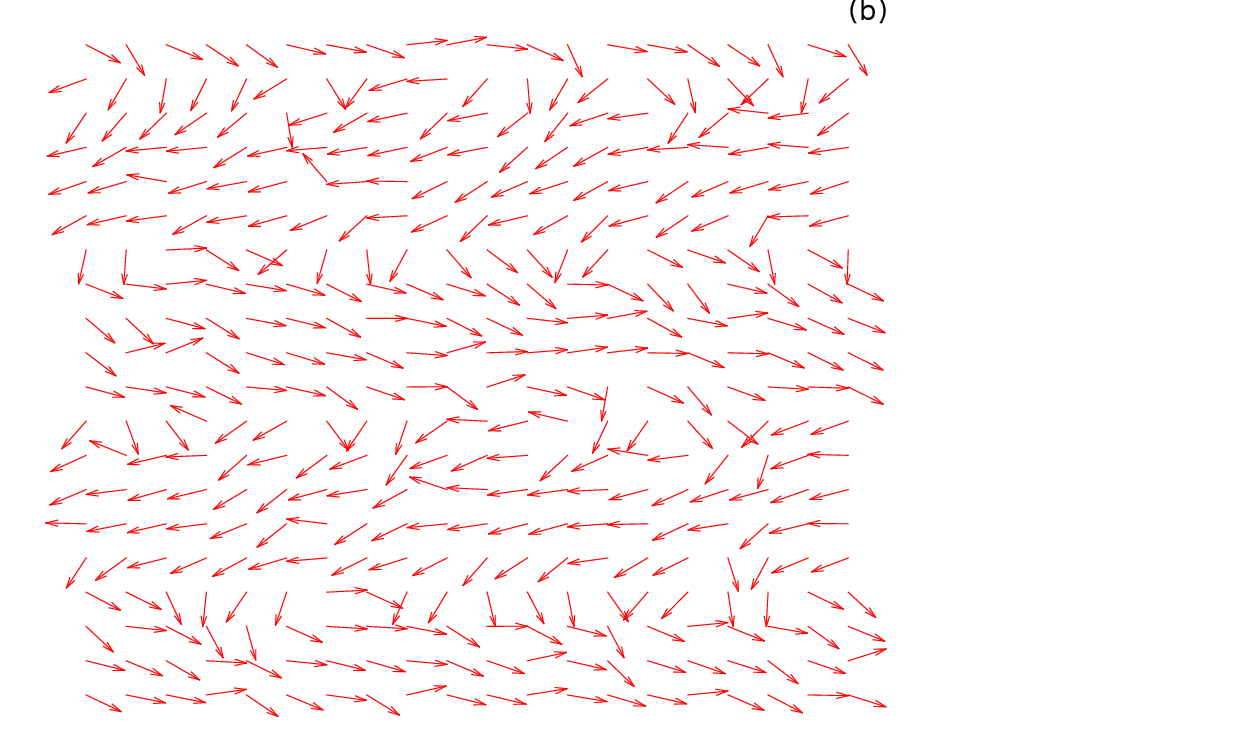}}
\\
\resizebox{8.5cm}{!}{\includegraphics[angle=0]{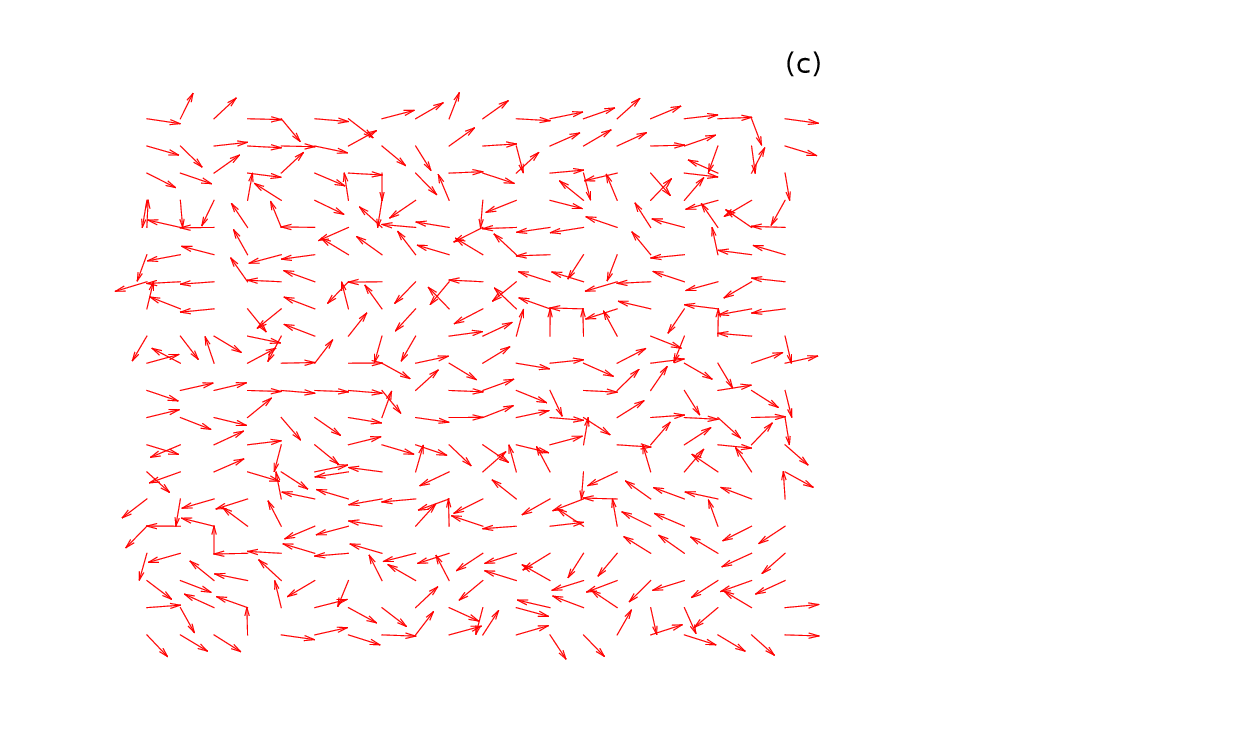}}
          \end{tabular}
\caption{Coherent propagation of driven spin-wave in the XZ plane (Y=10).
Wave propagates along the vertical direction.
(a) $t=1900$ MCSS and $T=0.4$, (b) $t=1930$ MCSS and $T=0.4$ and 
(c) $t=2000$ MCSS and $T=2.6$. Here, in all cases, $L=20$, $f=0.01$,
$\lambda=10$ and $H=3.0$.}
\label{fig:proplatt}
\end{center}
\end{figure}

The dynamical responses of the three dimensional XY ferromagnet, driven by 
linearly polarised
propagating magnetic field wave (described above) are studied. Starting
from a random initial spin configuration, the system is slowly (in the 
step of $\Delta T=0.02$ considered here) 
cooled down to achieve a nonequilibrium steady state. 
Depending on the values of $T$ and $H$ two distinct dynamical phases are
observed. In the low temperature, the coherent motion of bands of spins
oriented along a particular directions (on average), is found. One such
motion of spin bands are shown in Fig.~\ref{fig:proplatt}. 
The spin configuration of a XZ plane (Y=10) is shown here.
In the figure, the x-component of the spin lies along the horizontal axis
and y-component of the spin lies on the vertical axis.
Fig-\ref{fig:proplatt}(a) and Fig-\ref{fig:proplatt}(b) show
the configurations (for T=0.4 and H=3.0) of spins at two different 
instants (1900 and 1930 MCSS). The propagating modes
of spin bands, along the direction (upward here) of 
propagation of magnetic wave are clear. This is propagating spin wave mode,
observed in XY ferromagnet, driven by linearly polarised (along
horizontal direction) propagating (along the Z direction) magnetic field wave.
It may be noted here that magnetic wave propagates along the vertical direction
and the y-component of spin is shown along the same direction. The lattice
has dimensionality three 
(cubic system) whereas the spin has dimensionality two (XY model). This is the 
only way one can show the spin configuration of XZ plane in two dimensions. 
The low temperature spin configurations show that the x-component of spin is 
nearly zero, on an average. 
This is due to the response to the linearly (along x-direction) polarised 
propagating magnetic field wave.
However, the y-component (on average) of the spin is nonzero.
On the other hand, the
high temperature ($T=2.6$) spin configuration 
(at instant $t=2000$)
is completely random 
(or structureless) 
and has been shown in Fig-\ref{fig:proplatt}(c). In this case, 
both x and y components
of the spins are zero separately, on an average.

\begin{figure}[h]
\begin{center}
\begin{tabular}{c}
\resizebox{7cm}{!}{\includegraphics[angle=0]{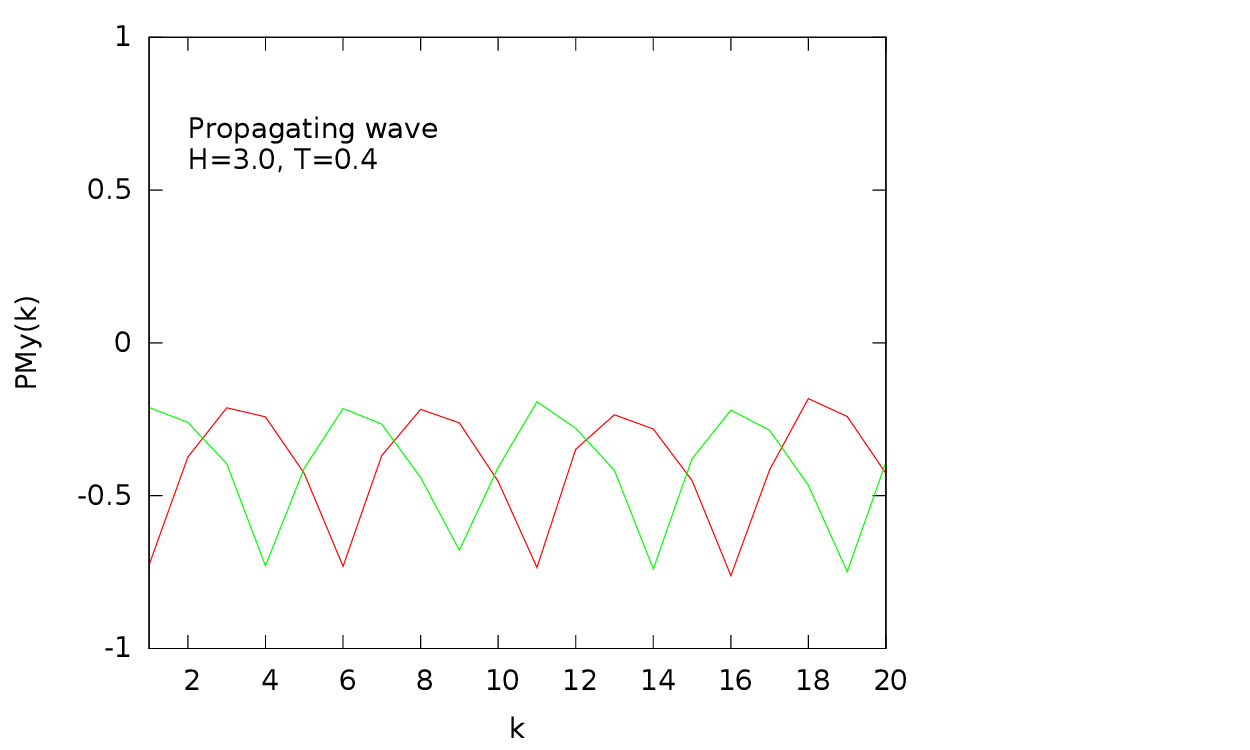}}

          \end{tabular}
\caption{The y-component of planar magnetisation PMy(k) plotted against k
(k-th XY plane). Different colors represent different time (a) t=1900 MCSS
(red line) and (b) t=1930 MCSS (green line). The propagating wave moves along
z direction.}
\label{fig:proppmyk}
\end{center}
\end{figure}

The propagating mode can also be visualised in another way. The wave
is propagating along the z direction. The y component of 
instantaneous planar magnetisation 
of any k-th XY plane is calculated as $PMy(k)={{1} \over {L^2}}
\sum s_y$,
where the sum is carried over all lattice sites of k-th XY plane. The planar
magnetisation (y-component) is plotted against k for two different times
and shown in Fig-\ref{fig:proppmyk}. From the figure the propagating mode is
evident.

\begin{figure}[h]
\begin{center}
\begin{tabular}{c}
\resizebox{7cm}{!}{\includegraphics[angle=0]{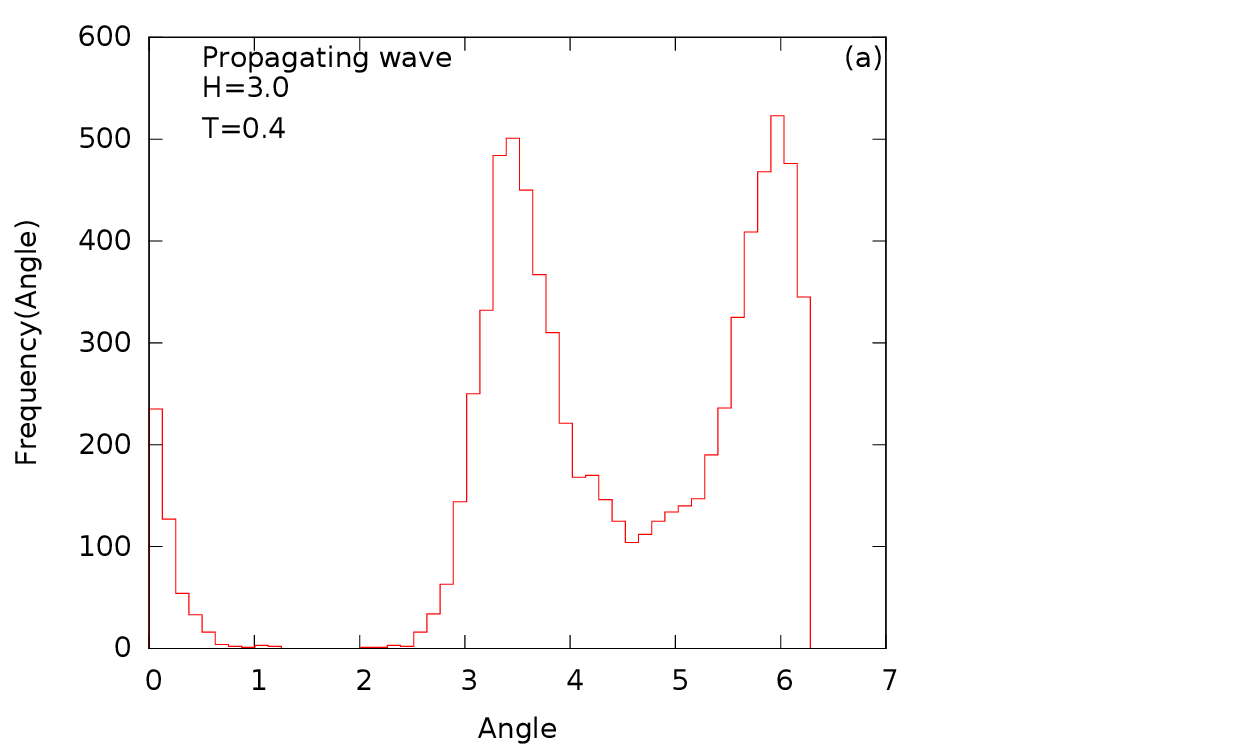}}
\\
\resizebox{7cm}{!}{\includegraphics[angle=0]{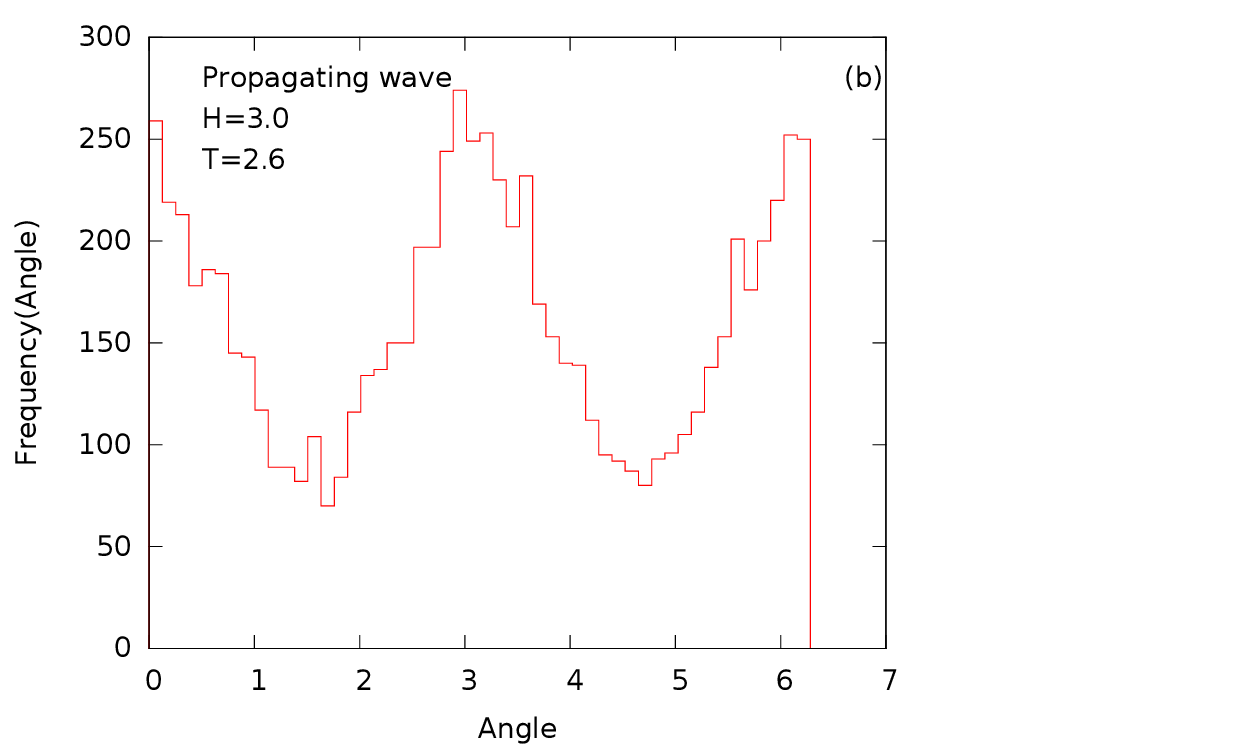}}
          \end{tabular}
\caption{Unnormalized distribution of angles of the spin vectors
for propagating waves.
(a) $t=2000$ MCSS and $T=0.4$, (b) $t=2000$ MCSS and $T=2.6$. 
Here, in all cases, $L=20$, $f=0.01$,
$\lambda=10$ and $H=3.0$.}
\label{fig:propangle}
\end{center}
\end{figure}

The dynamically stable spin structures, in the low temperature of the 
system in response to the propagating magnetic field wave are also observed
from the study of the statistical distribution of angles (of the spin). The
angle $\phi = {\rm tan^{-1}}({{s_y} \over {s_x}})$, where $s_x$and $s_y$ are
the values of x and y component of the spin respectively. The distribution
of $\phi$, over all spins ($L^3=8000$ here), are calculated at any instant
($t$=2000 here) for two different temperatures. One such distribution
(unnormalised), for $T=0.3$ is shown in Fig-\ref{fig:propangle}(a).
This distribution is trimodal. The three modes occur near $\phi=0$,
$\phi \simeq \pi$ (actually slightly above) and $\phi \simeq 2\pi$
(actually slightly below). 
The distribution vanishes at $\phi={{\pi}
\over {2}}$. The distribution gets significantly valuable 
(nonzero)
near $\phi \simeq
{{3\pi} \over {2}}$. This distribution of angle $\phi$ assures the 
existence of a net y component (here negative) 
with vanishingly small x component of spins. 
Other kind of distribution
of the angles are equally probable (for any other independent sample) which
will result the positive net y component.
The high temperature ($T=2.6$) structureless spin configuration is
justified by the distribution of angle ($\phi$). This is shown in
Fig-\ref{fig:propangle}(b), where the similar trimodal distribution
is observed with almost equal weightage of angles $\phi=
0, \pi$ and $2\pi$ leading to vanish the net x component of spin. 
Similarly, almost equal weightage of angles $\phi= {{\pi} \over {2}}, 
{{3\pi} \over {2}}$ compels net y component to vanish.

\begin{figure}[h]
\begin{center}
\begin{tabular}{c}
\resizebox{7cm}{!}{\includegraphics[angle=0]{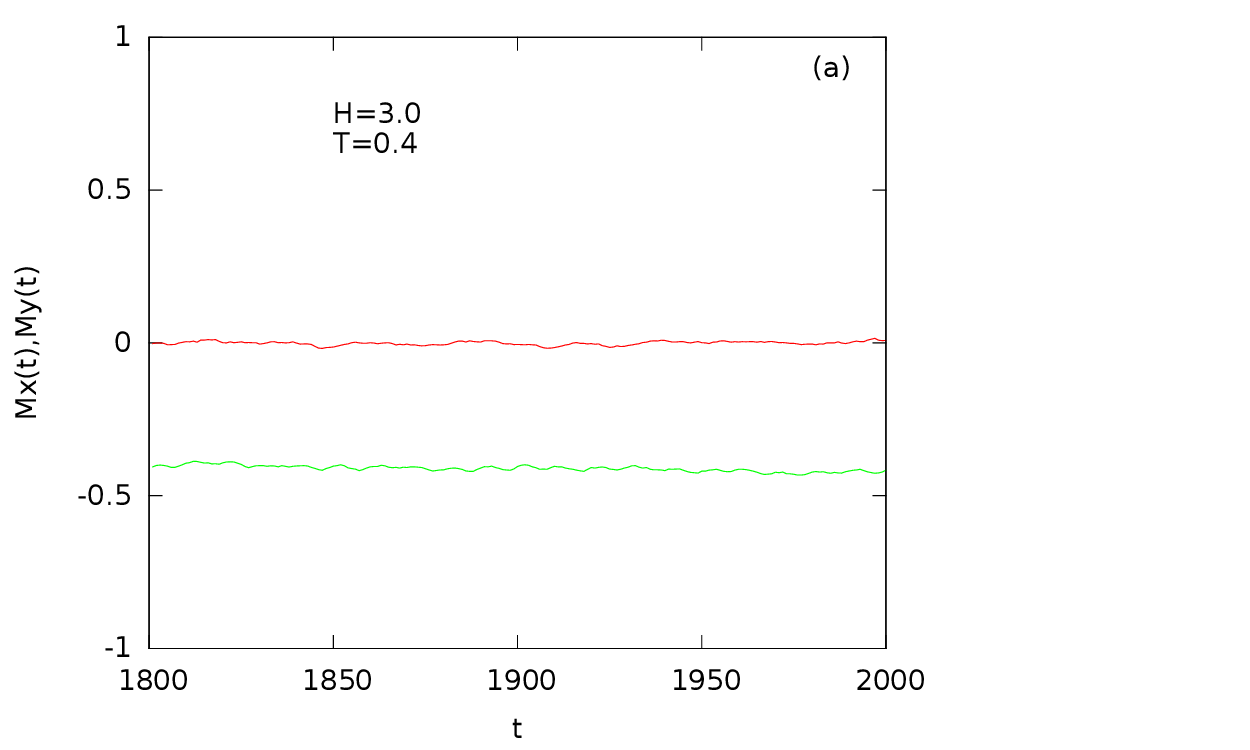}}
\\
\resizebox{7cm}{!}{\includegraphics[angle=0]{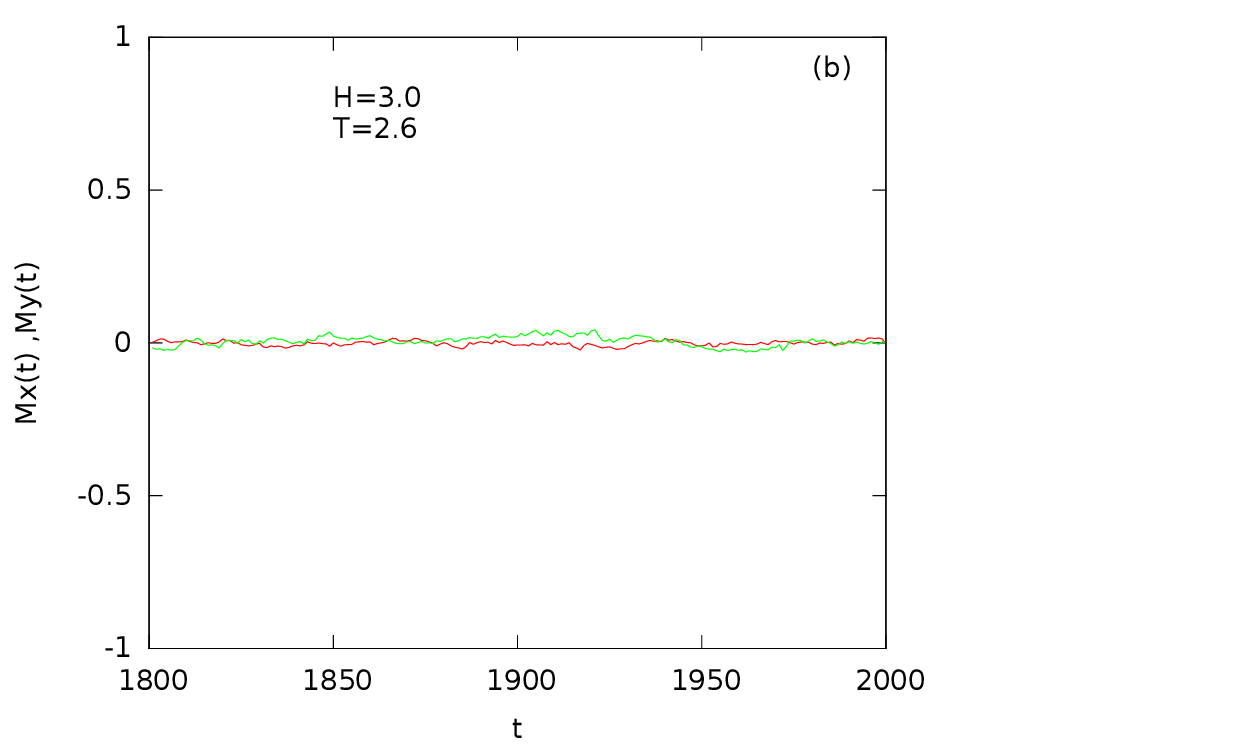}}
          \end{tabular}
\caption{Breaking of the dynamical symmetry in the case of propagating
wave. Plots of Mx(t)
(red line)  and My(t) (green line) as
functions of time (t) for different temperatures (a) T=0.4 (symmetry broken 
phase) and (b) T=2.6 (symmetric phase).
In both cases, H=3.0, f=0.01 and $\lambda=10$.}
\label{fig:propsym}
\end{center}
\end{figure}

The components (Mx(t) and My(t)) of instantaneous magnetisation are studied as 
functions of time and shown in Fig.~\ref{fig:propsym}. 
The x-component of instantaneous magnetisation,
Mx(t), is close to zero (apart from minor fluctuations) in 
the low temperature (T=0.4). However,
the y-component of this, shows a nonzero value 
(with some fluctuations). On the other
hand, in the high temperature (T=2.6) both vanish
(fluctuate around zero). The system undergoes a {\it partial 
breaking of dynamical symmetry} 
(My(t) only) as it cooled down from high temperature. The 
existence of the
partially dynamic symmetry broken phase, in the low temperature regime,
 is also evident from the distribution
of angles discussed above.

\begin{figure}[h]
\begin{center}
\begin{tabular}{c}
\resizebox{7cm}{!}{\includegraphics[angle=0]{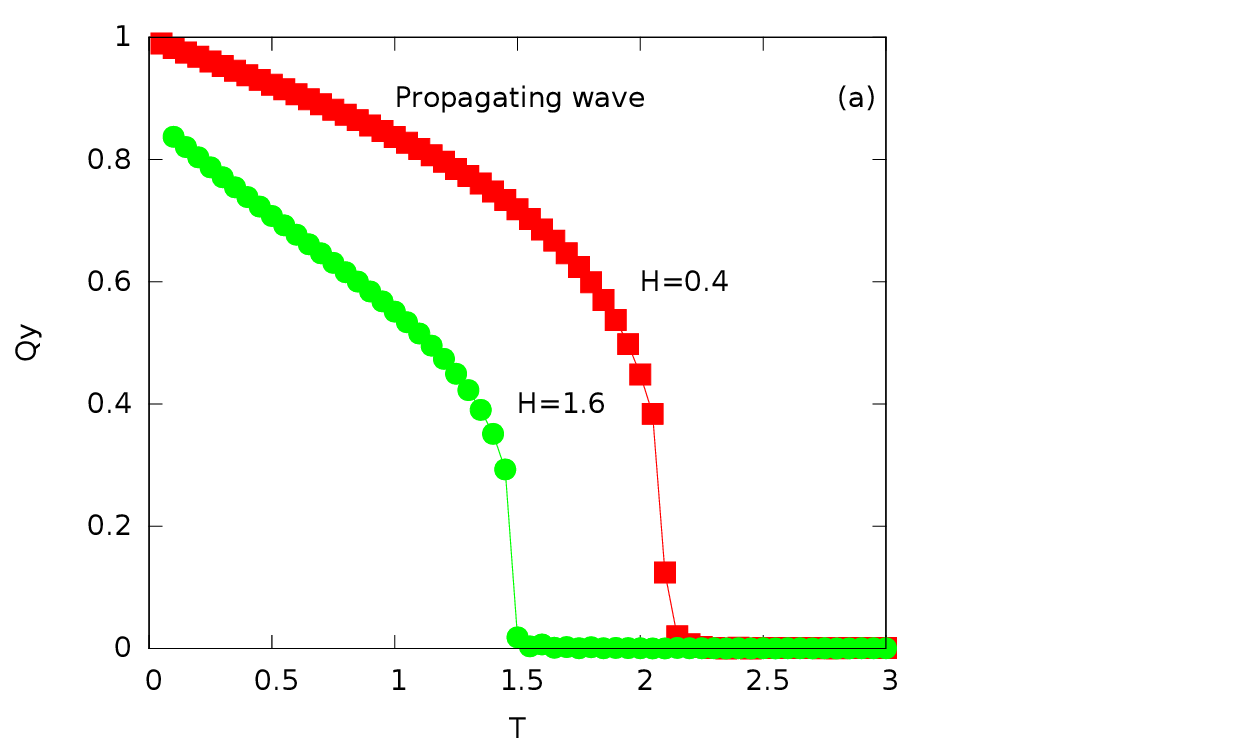}}
\\
\resizebox{7cm}{!}{\includegraphics[angle=0]{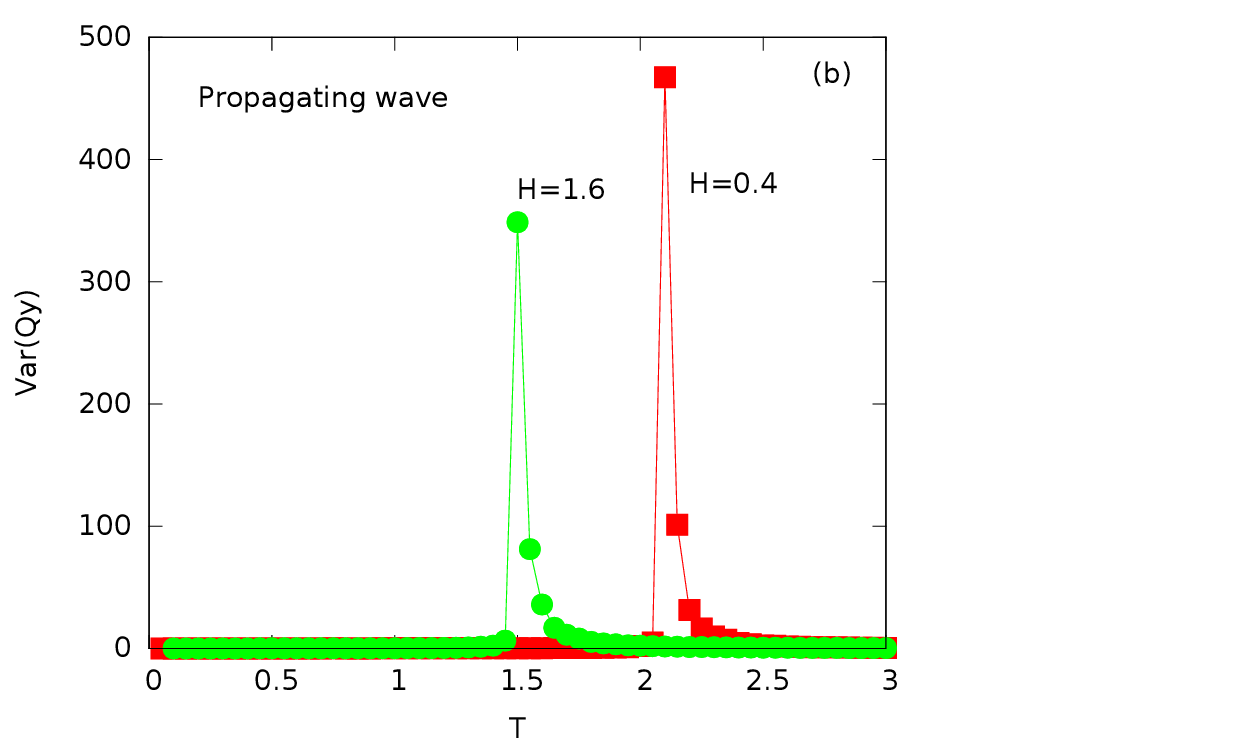}}
\\
\resizebox{7cm}{!}{\includegraphics[angle=0]{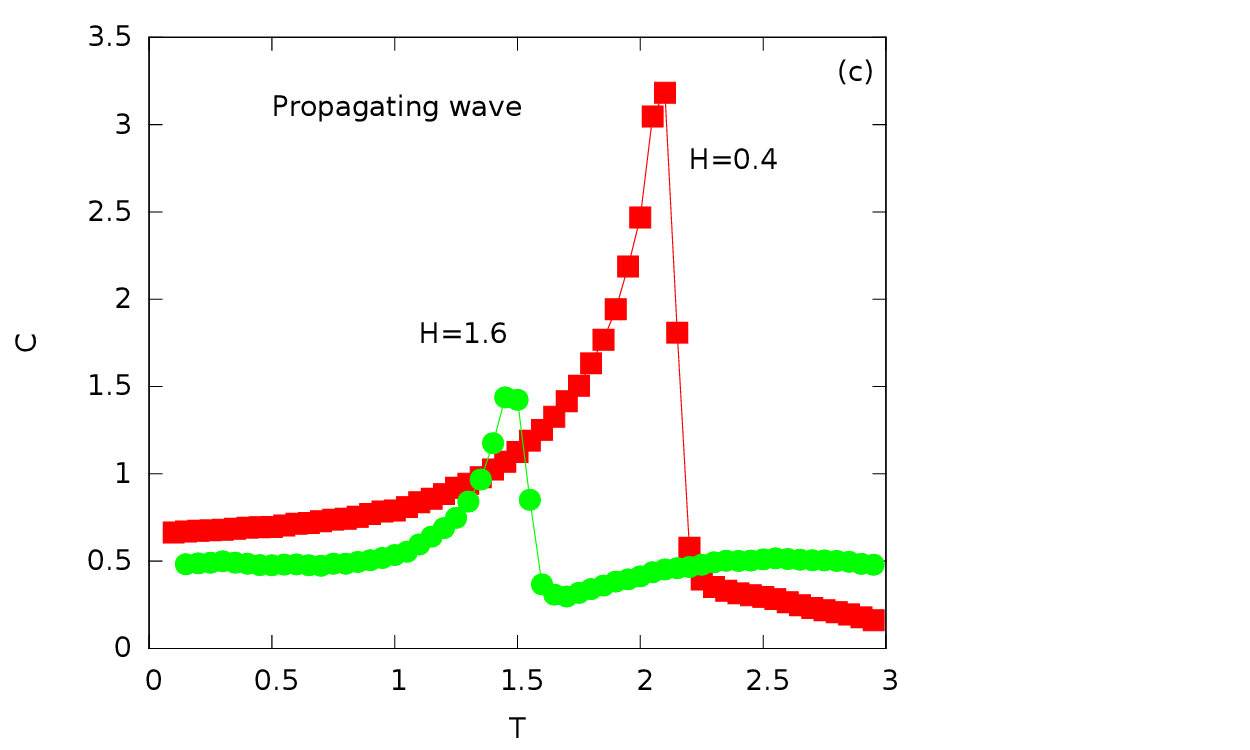}}
          \end{tabular}
\caption{$Qy$, $Var(Qy)$ and $C$ are plotted as functions of the temperature ($T$)
for two different values of the amplitude ($H$) of propagating field wave.
Here, in all cases, $L=20$, $f=0.01$,
$\lambda=10$.}
\label{fig:propall}
\end{center}
\end{figure}

From the usual definition of dynamic order parameter Qx and Qy, it is clear that as the
system is cooled down, Qy gets a nonzero value (corresponding to dynamically symmetry
broken ordered phase) from Qy=0 (corresponding to dynamically symmetric 
disordered phase). Needless
to say that Qx=0 always. Fig.~\ref{fig:propall}(a) shows the variation of Qy as a function
of temperature. There exists a critical temperature, below which Qy$\neq 0$, is so called dynamic
transition temperature. This dynamic transition temperature is found to decrease as the 
amplitude (H) of the propagating field increases. The variance of Qy, i.e., Var(Qy) is 
found to become sharply peaked at the transition
point (Fig.~\ref{fig:propall}(b)). From this diagram, the dynamic transition temperature, is
determined and found to decrease as the amplitude of the field (H) increases.
Interestingly, the dynamic transition point is also indicated by the 
temperature, for which the dynamic
specific heat C (studied as function of 
temperature T), gets sharply peaked (Fig.~\ref{fig:propall}(c)).

\begin{figure}[h]
\begin{center}
\begin{tabular}{c}
\resizebox{7cm}{!}{\includegraphics[angle=0]{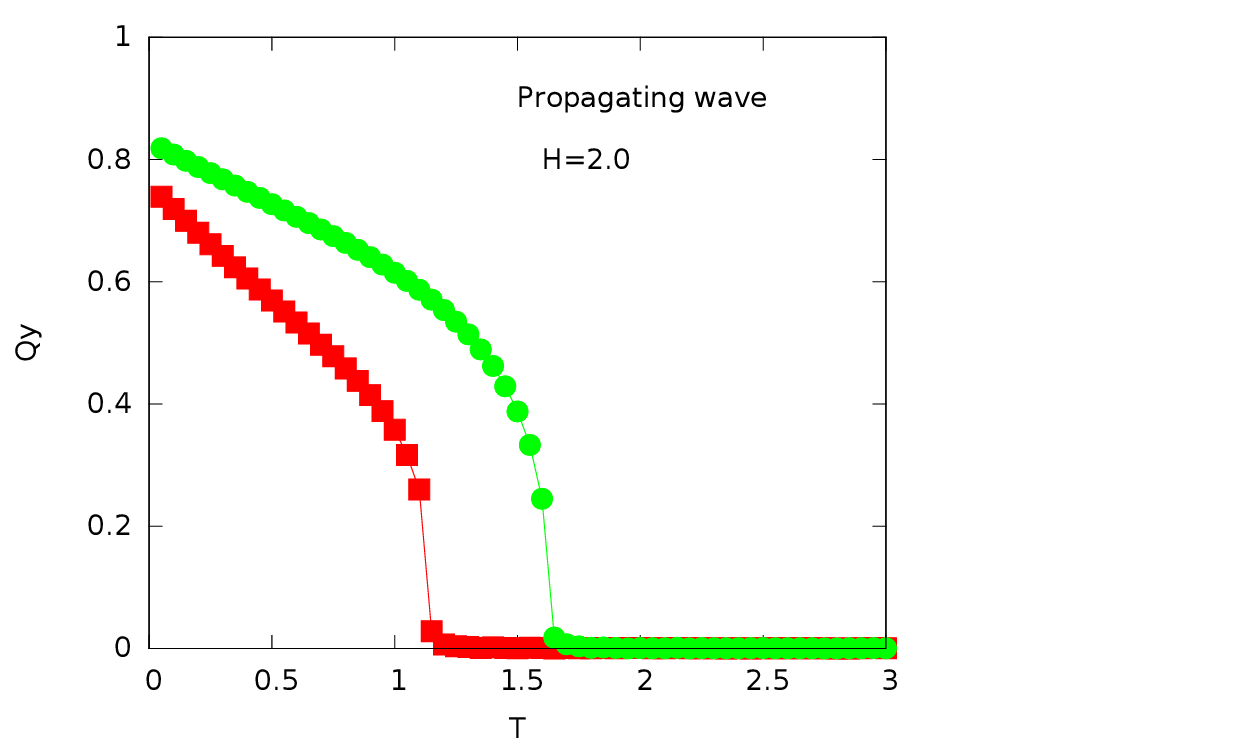}}
\\
\resizebox{7cm}{!}{\includegraphics[angle=0]{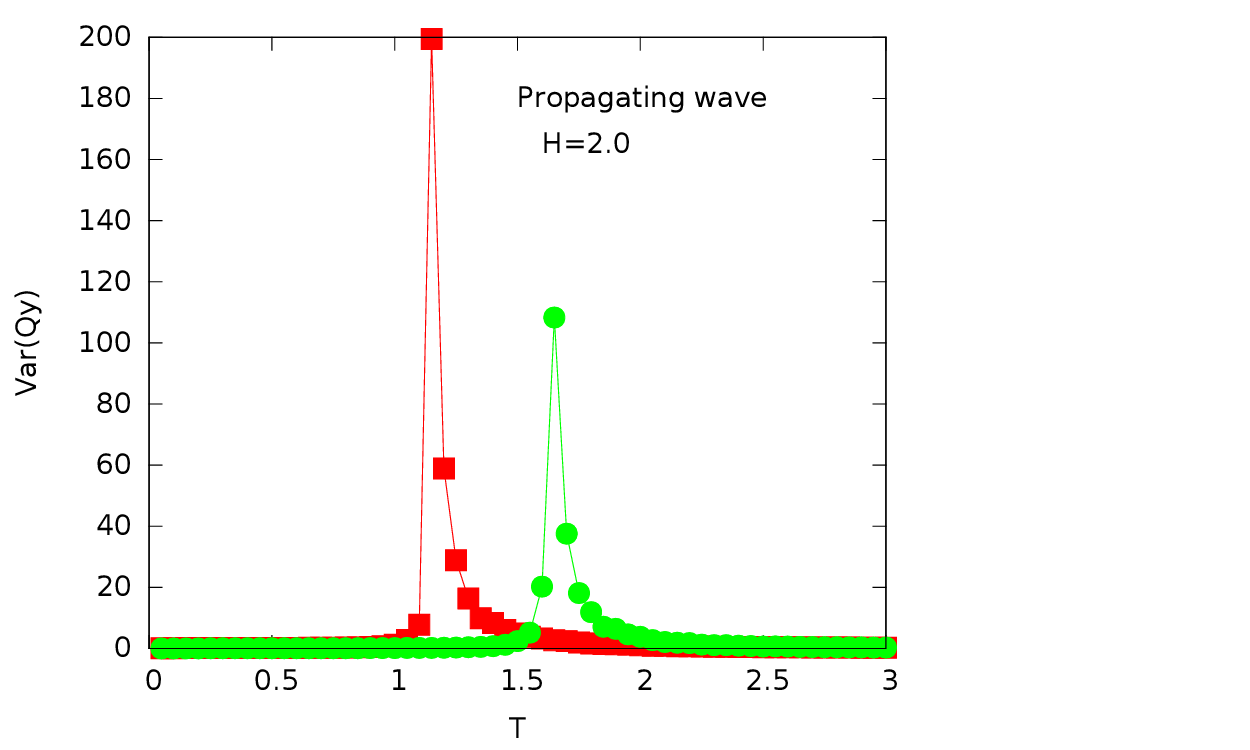}}
          \end{tabular}
\caption{Qy and Var(Qy) are plotted as functions of the temperature ($T$)
for two different values of the wavelengths ($\lambda$) of propagating field wave.
Red square ($\lambda=20$) and Green bullet ($\lambda=5$).
Here, in all cases, $L=20$, $f=0.01$ and $H=2.0$}
\label{fig:prop2wl}
\end{center}
\end{figure}

The dynamic transition temperatures are found to vary with wavelength of the propagating
magnetic wave (for fixed value of amplitude). This is shown in Fig.~\ref{fig:prop2wl}. The
dynamic order parameter Qy and its variance Var(Qy) are studied as functions of temperature
(T) for a two different values 
($\lambda=$ 20 and 5)
of the wavelength of the propagating magnetic field wave (but
with fixed amplitude H=2.0). From the figure it is clear that, the system transits (order
- disorder) at lower temperature (fixed $H$) for longer waves of propagating 
magnetic field wave.

\begin{figure}[h]
\begin{center}
\begin{tabular}{c}
        \resizebox{8cm}{!}{\includegraphics[angle=0]{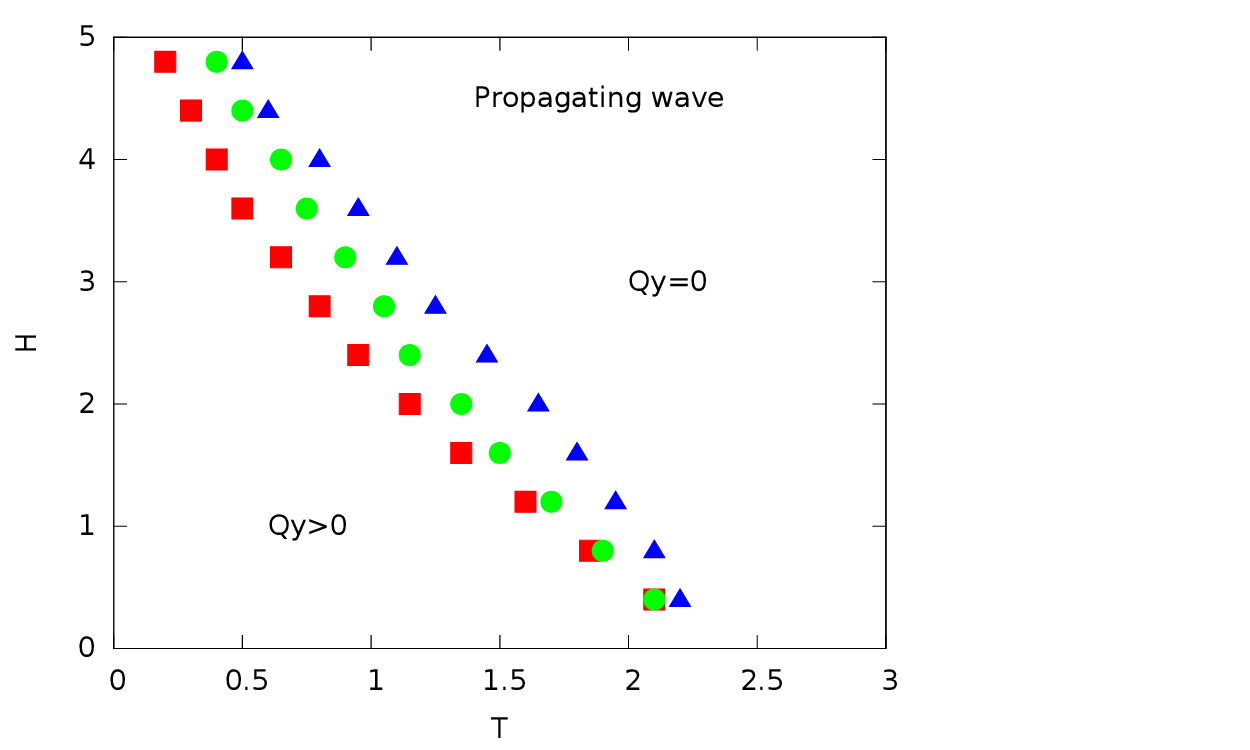}}
        
          \end{tabular}
\caption{The phase diagrams for {\it propagating} wave. Different symbols correspond
to the different values of wavelength ($\lambda$). (Red square) $\lambda=20$, 
(Green bullet) $\lambda=10$ and (Blue triangle) $\lambda = 5$.}
\label{fig:propphase}
\end{center}
\end{figure}

The dependences of the dynamic transition temperature, 
on the field amplitude and the wavelength
of the propagating magnetic field wave, have been represented in a comprehensive manner in the
phase diagram shown in Fig.~\ref{fig:propphase}. 
The phase boundaries are obtained for three different values of the wavelength
($\lambda=$ 20, 10 and 5)
of the propagating magnetic wave.
The phase boundary is found to shrink (towards low
temperature and low field) inward for longer waves of the propagating 
magnetic field wave.

\begin{figure}[h]
\begin{center}
\begin{tabular}{c}
        \resizebox{8cm}{!}{\includegraphics[angle=0]{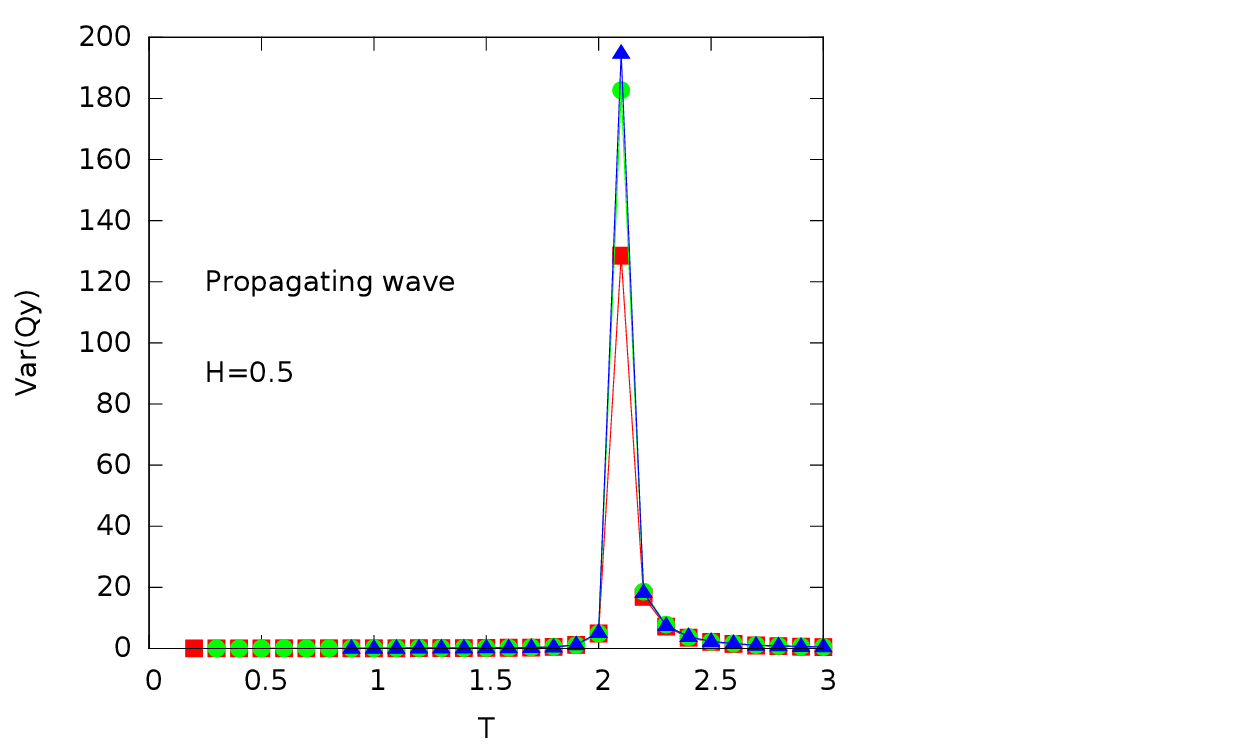}}
        
          \end{tabular}
\caption{The size dependence of the transition temperature in the case of {\it propagating} wave. Different symbols correspond
to the different values of system size ($L$). (Red square) $L=20$, 
(Green bullet) $L=30$ and (Blue triangle) $L=40$. The wavelength of the propagating wave $\lambda=10$ lattice unit in all three cases.}
\label{fig:fss-pw}
\end{center}
\end{figure}

The detailed finite size analysis, in the system of spins having continuous symmetry,
 is a huge computational task. One has to
determine the critical temperature from the intersection of the fourth order
Binder cumulant plotted as functions of temperature for different system sizes. 
Here, within the limited computational facilities and time,
the transition temperatures are calculated crudely from the peak positions of
$Var(Qy)$ (studied as function of temperature) for three different values of
$L$ ($L=20,30$ and 40). For all values of $L$, the wavelength $\lambda$ was kept 
fixed ($\lambda=10$). Within the used accuracy ($\Delta T=0.02$ in this
simulation), it is observed that the positions of the peaks (for different $L$)
remain same (Fig-\ref{fig:fss-pw}). 
However, the peak height increases as $L$
increases showing the growth of critical correlation (see Fig-2 of ref\cite{sides}).
This is a peculiar result which shows that (at least from this simulation) the finite
size effects do not have any strong dependence on the wavelength $\lambda$. However,
further detail investigation is required to have strongly conclusive results.

\vskip 0.5cm

\noindent {\it (b) Standing wave:} 

\begin{figure}[h]
\begin{center}
\begin{tabular}{c}
\resizebox{7cm}{!}{\includegraphics[angle=0]{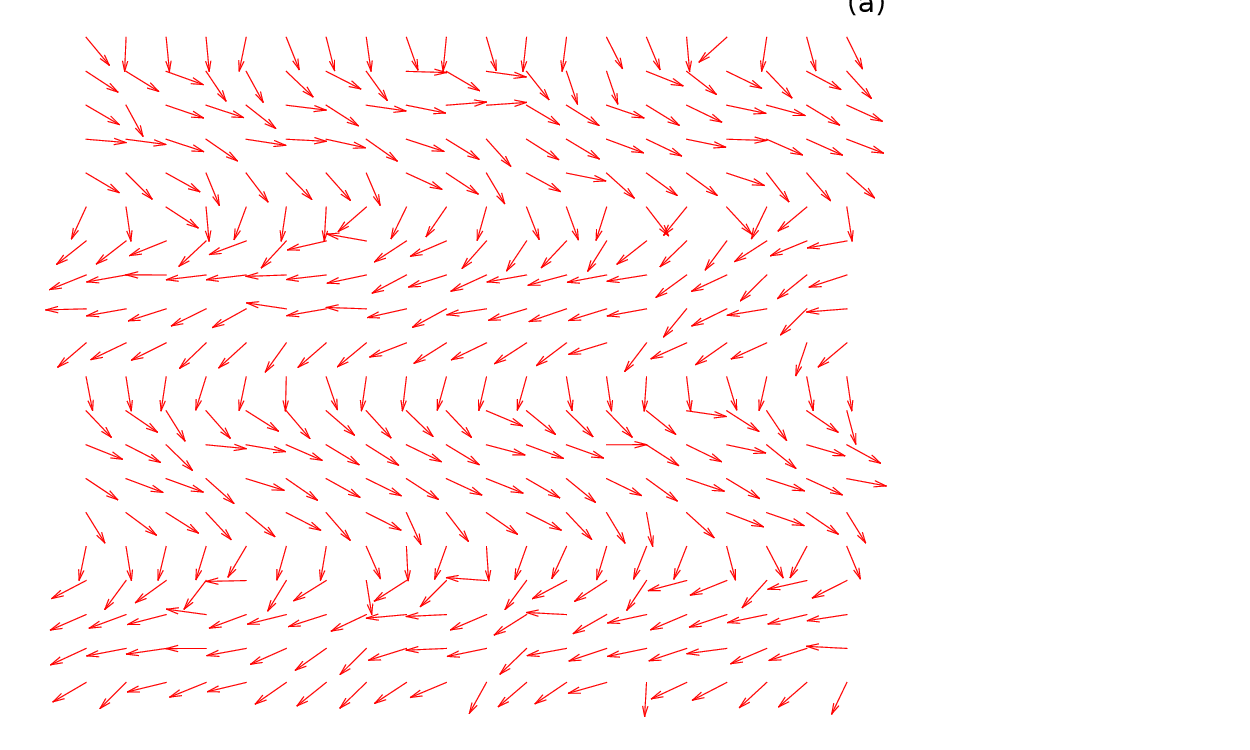}}

\\
\resizebox{7cm}{!}{\includegraphics[angle=0]{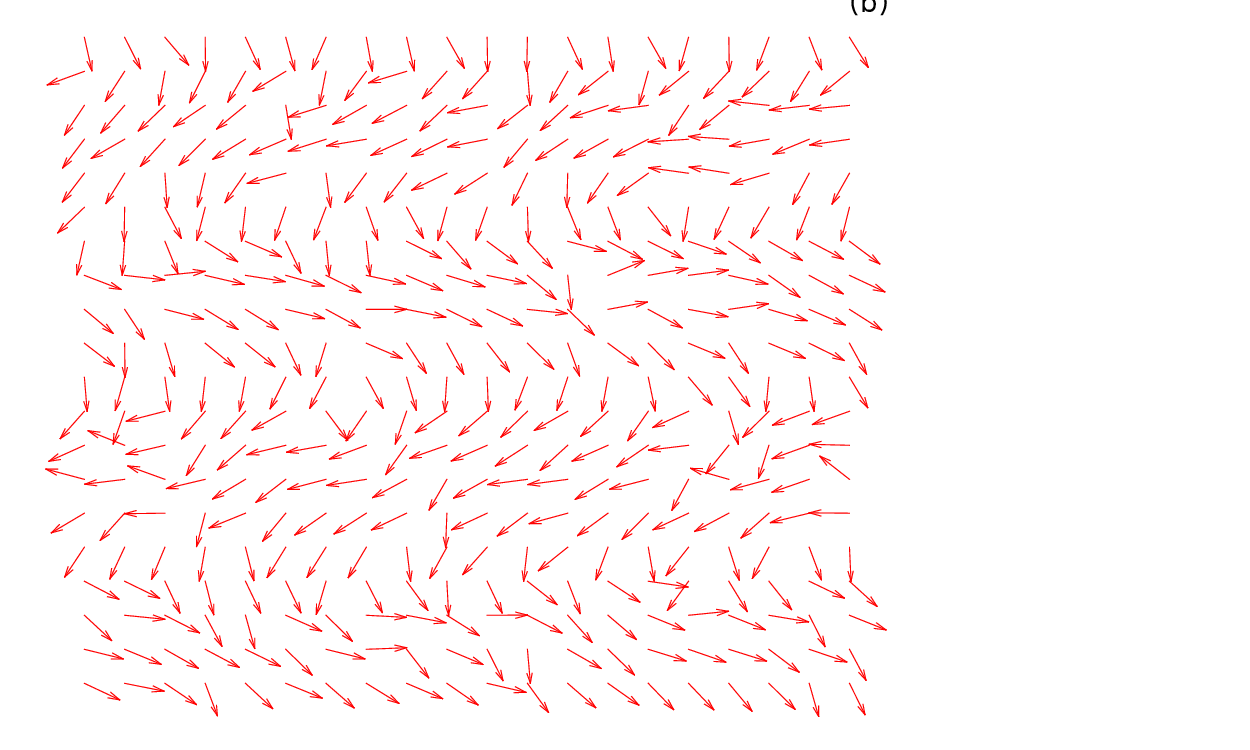}}

\\
\resizebox{7cm}{!}{\includegraphics[angle=0]{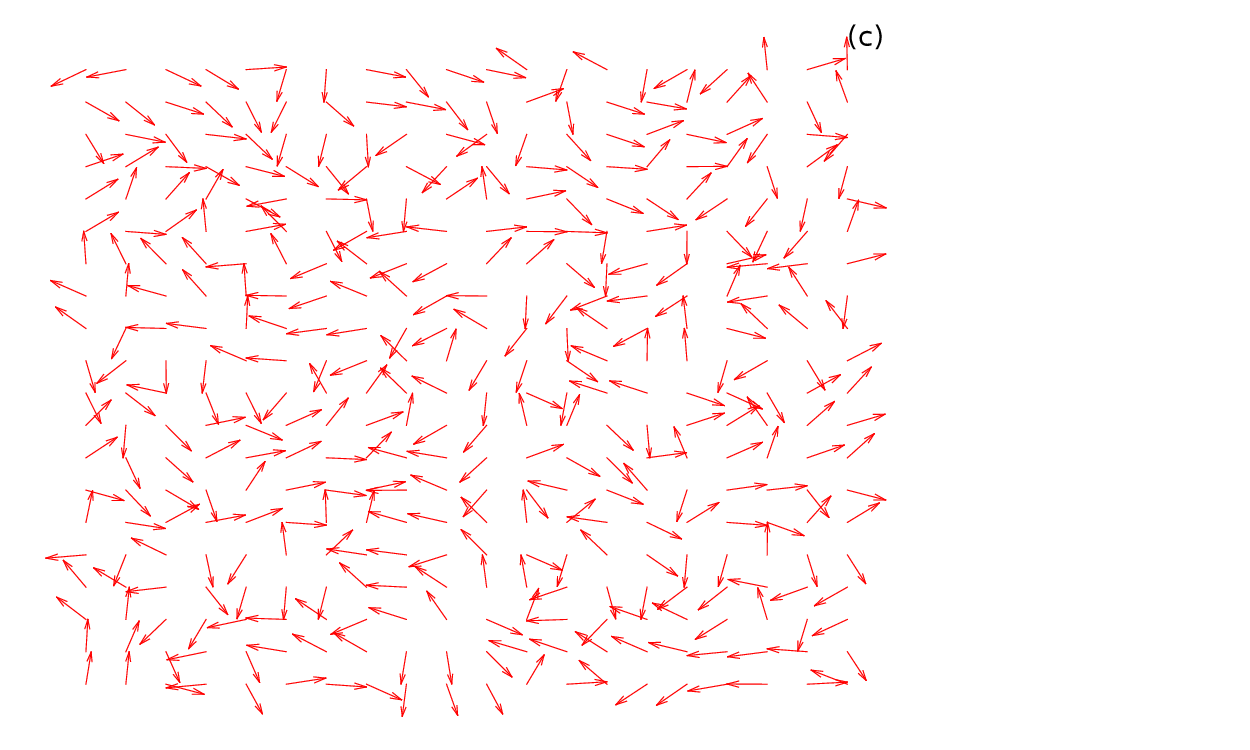}}

          \end{tabular}
\caption{Standing mode of driven spin-wave in the XZ plane (Y=10).
(a) $t=1900$ MCSS and $T=0.4$, (b) $t=1930$ MCSS and $T=0.4$ and 
(c) $t=2000$ MCSS and $T=2.6$. Here, in all cases, $L=20$, $f=0.01$,
$\lambda=10$ and $H=3.0$.}
\label{fig:standlatt}
\end{center}
\end{figure}

Now let us see what happens to the XY ferromagnet if it is driven by standing
magnetic field wave. The standing magnetic field wave is extended along the
z axis and the polarisation of the field is linear (along the x axis). Here
also, two distinct dynamical modes are observed depending upon the values of
$T$ and $H$. For a fixed value of $H=3.0$ the low temperature ($T=0.4$) and
high temperature ($T=2.6$) spin configurations of any XZ plane (Y=10 here)
are shown in Fig-\ref{fig:standlatt}. In the low temperature, the standing
spin wave modes are observed and shown in Fig-\ref{fig:standlatt}.
Unlike the case of propagating spin wave mode, here the spin bands are formed
but not changing their positions in time. 
This indeed reveals the standing spin wave mode.
Fig-\ref{fig:standlatt}(a) and
Fig-\ref{fig:standlatt}(b) shows the spin configurations for two different
instants ($t=1900$ and $t=1930$ MCSS). The spin bands are found not to
change their positions in time. 
However, in the high temperature, the structureless
or random spin configration was observed (shown in Fig-\ref{fig:standlatt}(c)).

\begin{figure}[h]
\begin{center}
\begin{tabular}{c}
\resizebox{7cm}{!}{\includegraphics[angle=0]{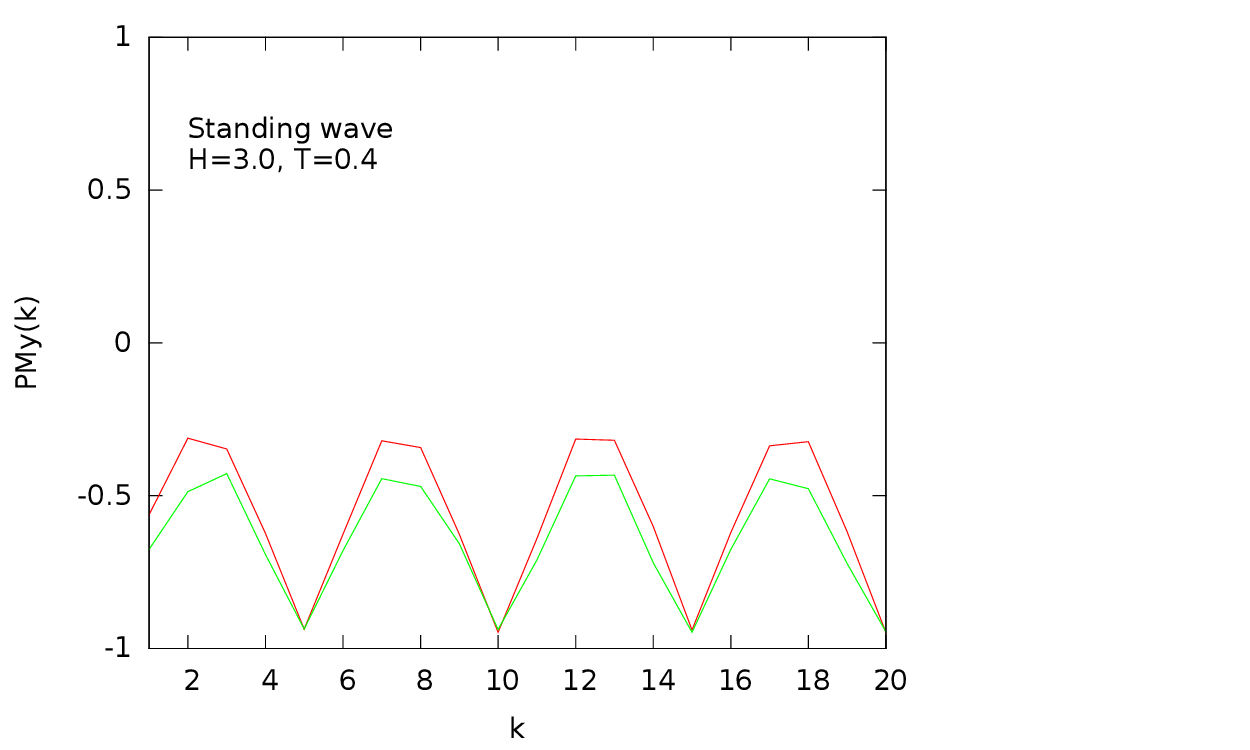}}

          \end{tabular}
\caption{The y-component of planar magnetisation PMy(k) plotted against k
(k-th XY plane). Different colors represent different time (a) t=1900 MCSS
(red line) and (b) t=1930 MCSS (green line). The standing wave extends along
z direction.}
\label{fig:standpmyk}
\end{center}
\end{figure}

The standing mode of driven spin wave in XY ferromagnet, can also be 
visualised in a different way. The y component of planar magnetisation
$PMy(k)$ of k-th XY plane are plotted against k and shown in 
Fig-\ref{fig:standpmyk} for two different times. From the diagram, it is
clear that the profiles ($PMy(k)$ versus $k$) do not change their 
relative positions
in time (unlike the case of propagating mode shown in Fig-\ref{fig:proppmyk}).

\begin{figure}[h]
\begin{center}
\begin{tabular}{c}
\resizebox{7cm}{!}{\includegraphics[angle=0]{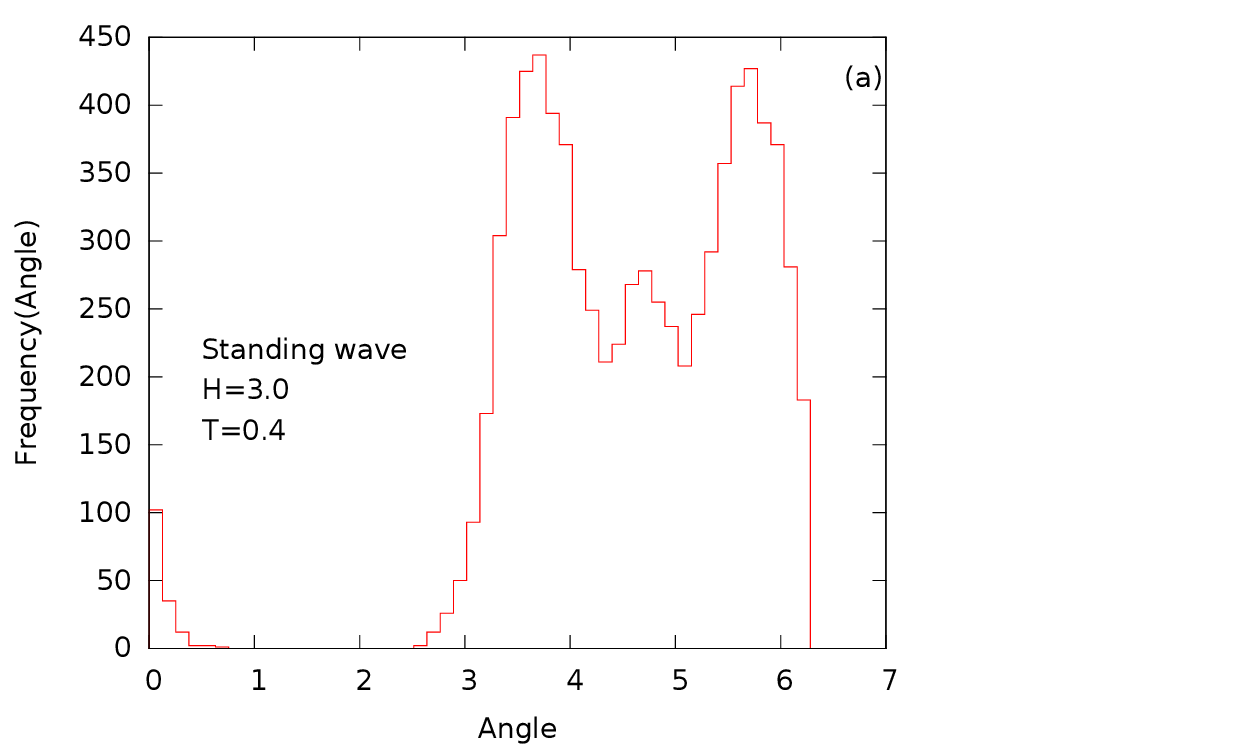}}
\\
\resizebox{7cm}{!}{\includegraphics[angle=0]{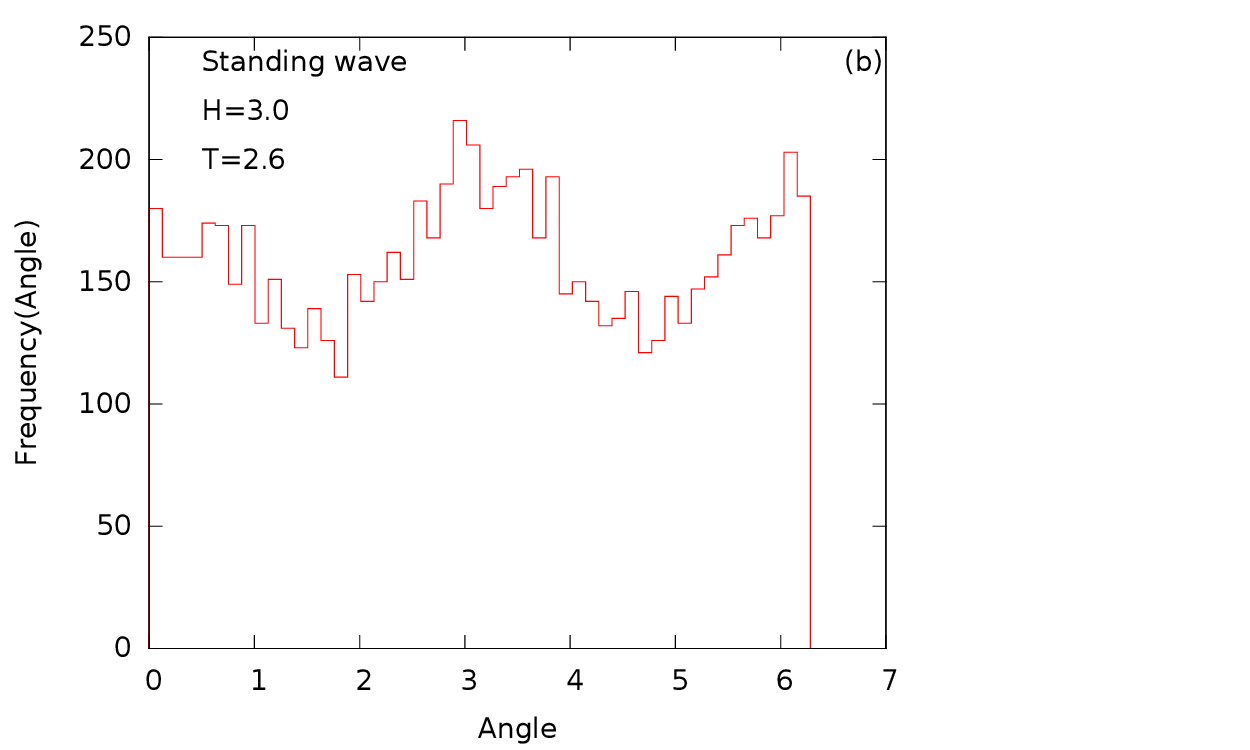}}
          \end{tabular}
\caption{Unnormalized distribution of angles of the spin vectors
for standing magnetic waves.
(a) $t=2000$ MCSS and $T=0.4$, (b) $t=2000$ MCSS and $T=2.6$. 
Here, in all cases, $L=20$, $f=0.01$,
$\lambda=10$ and $H=3.0$.}
\label{fig:standangle}
\end{center}
\end{figure}

The existence of the low temperature ordered (or structured) 
standing spin wave mode and
the high temperature disordered (or structureless) dynamical states can
be realised through the statistical distributions of the angles of the
spin vectors. The statistical distribution of the angle (of spin vector) $\phi$
is studied and shown in Fig-\ref{fig:standangle}. The low temperature 
unnormalised distribution is shown in Fig-\ref{fig:standangle}(a). Here,
the distribution is tetramodal (having four maxima). 
The modes (maxima) of the distributions are found at:
$\phi=0$, $\phi \simeq \pi$ (actually slightly above), 
$\phi \simeq {{3\pi} \over {2}}$ and
$\phi \simeq 2\pi$ (actually slightly below).
It may be noted that, there was no peak in the distribution of $\phi$,
near ${{3\pi} \over {2}}$, in the case of propagating wave 
(compare with Fig-\ref{fig:propangle}(a)).
The angle $\phi$ dislikes to accept
any value near ${{\pi} \over {2}}$. This peculiar kind of distribution  
of angle $\phi$ leads to net (nonzero) y component (negative here)
 of magnetisation. 
Here also, other kind of distribution
of the angles are equally probable (for any other independent sample) which
will result the positive net y component.
The x component of magnetisation is zero on an average.
This is
characterised as dynamically ordered phase. On the other hand, the high 
temperature distribution of angle $\phi$ 
(shown in Fig-\ref{fig:standangle}(b)) is almost symmetric and trimodal
(having three maxima of almost equal height) around $\phi=0$, $\phi \simeq \pi$
 and $\phi \simeq
2\pi$. Needless to mention that this 
would lead to a vanishing net y component
of magnetisation. The net x component of magnetisation is zero, on an average.
This state is characterised as dynamically disordered phase.
Like the case of propagating wave (already mentioned above), the ordered
phase is a symmetry broken phase and the disordered one is symmetric (in
all directions) phase. The partial (in y component only) 
symmetry broken ordered phase is observed in the case
of standing wave. 

\begin{figure}[h]
\begin{center}
\begin{tabular}{c}
\resizebox{7cm}{!}{\includegraphics[angle=0]{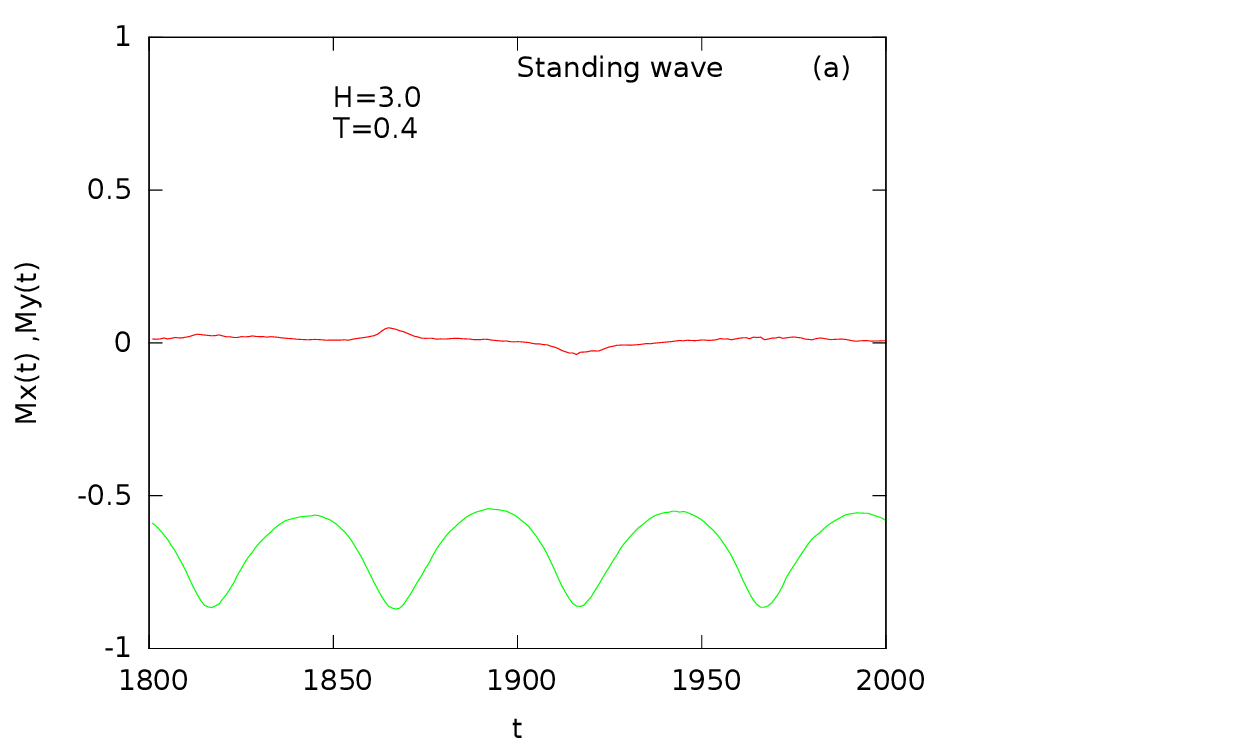}}

\\
\resizebox{7cm}{!}{\includegraphics[angle=0]{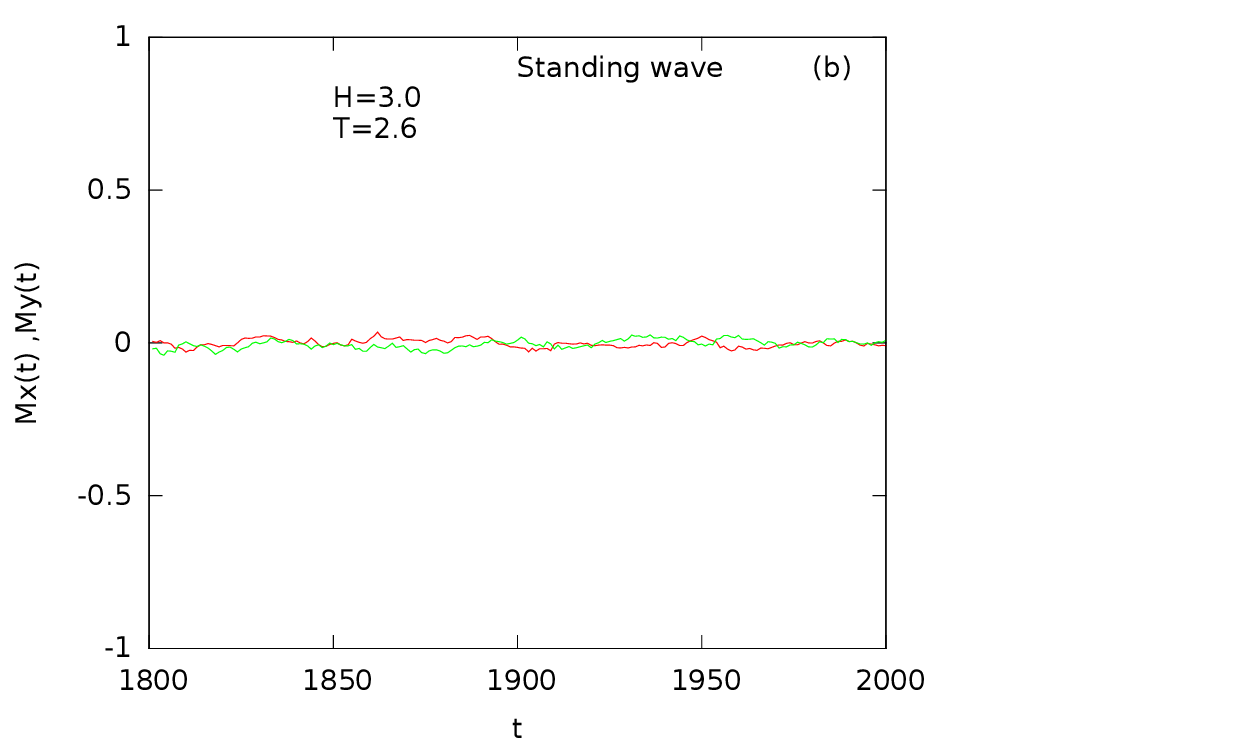}}
          \end{tabular}
\caption{Breaking of the dynamical symmetry in the case of standing wave. Plots of Mx(t)
(red line)  and My(t) (green line) as
functions of time (t) for different temperatures (a) T=0.4 
(symmetry broken phase) and (b) T=2.6 (symmetric phase).
In both cases, H=3.0, f=0.01 and $\lambda=10$.}
\label{fig:standsym}
\end{center}
\end{figure}

The dynamical symmetry breaking is clearly visible in the study of the time
dependences of the components of magnetisation.  
The instantaneous components 
($Mx(t)$ and $My(t)$)
of magnetisation are studied as functions of time and shown in 
Fig-\ref{fig:standsym}. The high temperature ($T=2.6$) study 
(in Fig-\ref{fig:standsym}(a)) shows the existence of a dynamically 
symmetric phase where both $Mx(t)$ and $My(t)$ varies almost symmetrically
around zero line. On the other hand, the dynamical
symmetry of the low temperature ($T=0.4$) phase
is partially (y component only) broken
(Fig-\ref{fig:standsym}(b)). In this case,
only the y component $My(t)$ varies asymmetrically (about zero line). So, a
symmetry breaking dynamic (or nonequilibrium) transition is expected in
the driven (by standing magnetic wave) XY ferromagnet. It may be noted that
in both phases, the time average x component of magnetisation over the full
cycle of the standing magnetic wave is zero.

\begin{figure}[h]
\begin{center}
\begin{tabular}{c}
\resizebox{7cm}{!}{\includegraphics[angle=0]{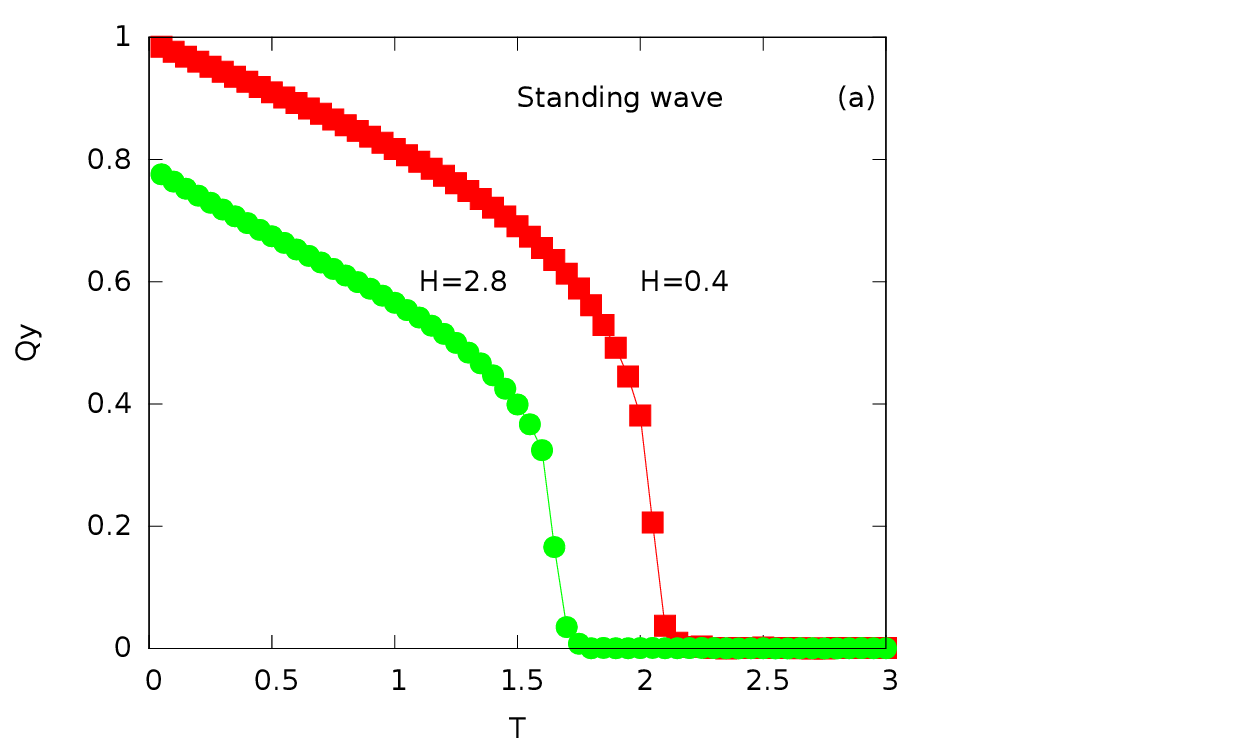}}

\\
\resizebox{7cm}{!}{\includegraphics[angle=0]{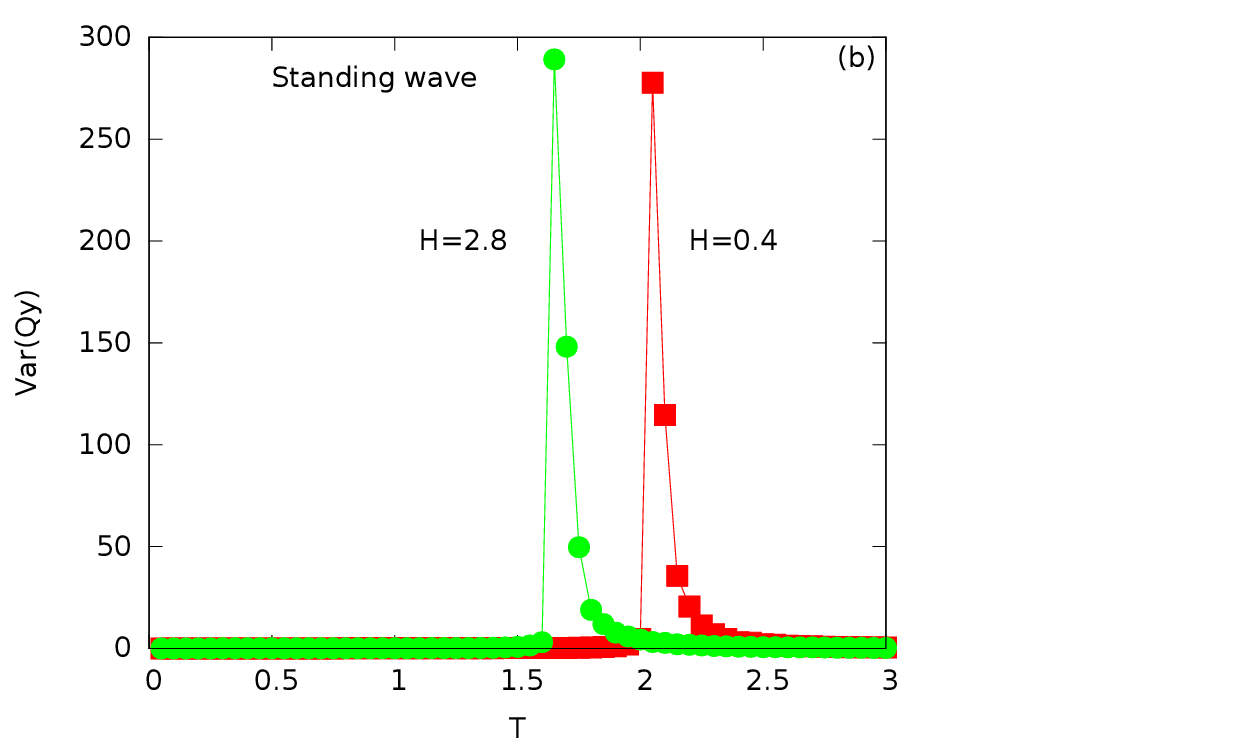}}

\\
\resizebox{7cm}{!}{\includegraphics[angle=0]{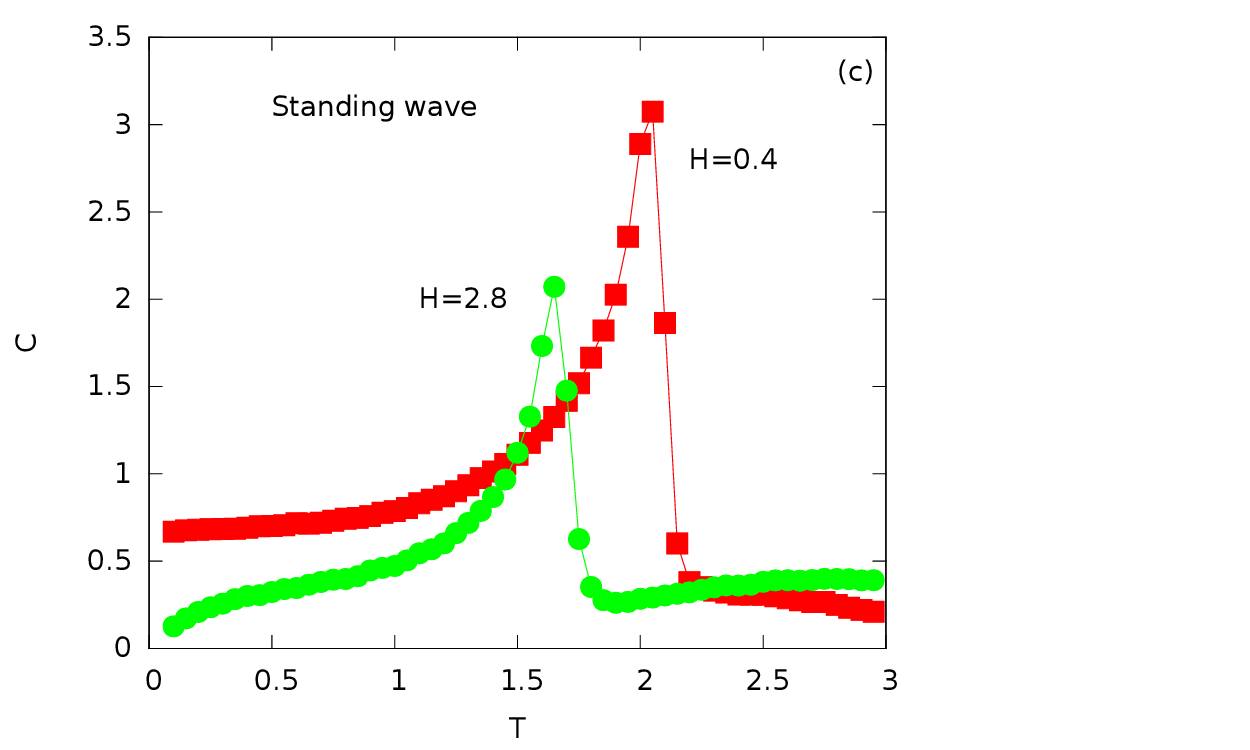}}
          \end{tabular}
\caption{Qy, Var(Qy) and C are plotted as functions of the temperature ($T$)
for two different values of the amplitude ($H$) of standing magnetic field wave.
Here, in all cases, $L=20$, $f=0.01$,
$\lambda=10$.}
\label{fig:standall}
\end{center}
\end{figure}

To study the symmetry breaking dynamic phase transition in driven XY
ferromagnet, the temperature dependences of $Qy$, $Var(Qy)$ and
$C$ are studied and shown in Fig-\ref{fig:standall}. For a fixed value of $H$,
the $Qy$ takes a nonzero value 
near a transition temperature
as the system is cooled down from a high
temperature. This transition temperature is found to decrease as the value of
$H$ is increased. This is shown in Fig-\ref{fig:standall}(a). The $Var(Qy)$
and $C$ show sharp peaks near the transition temperatures. They are shown in
Fig-\ref{fig:standall}(b) and Fig-\ref{fig:standall}(c). The peaks determine
the transition temperatures. From all these studies it is clear that the
transitions occur at lower temperatures for higher values of the field
amplitudes ($H$).

\begin{figure}[h]
\begin{center}
\begin{tabular}{c}
        \resizebox{7cm}{!}{\includegraphics[angle=0]{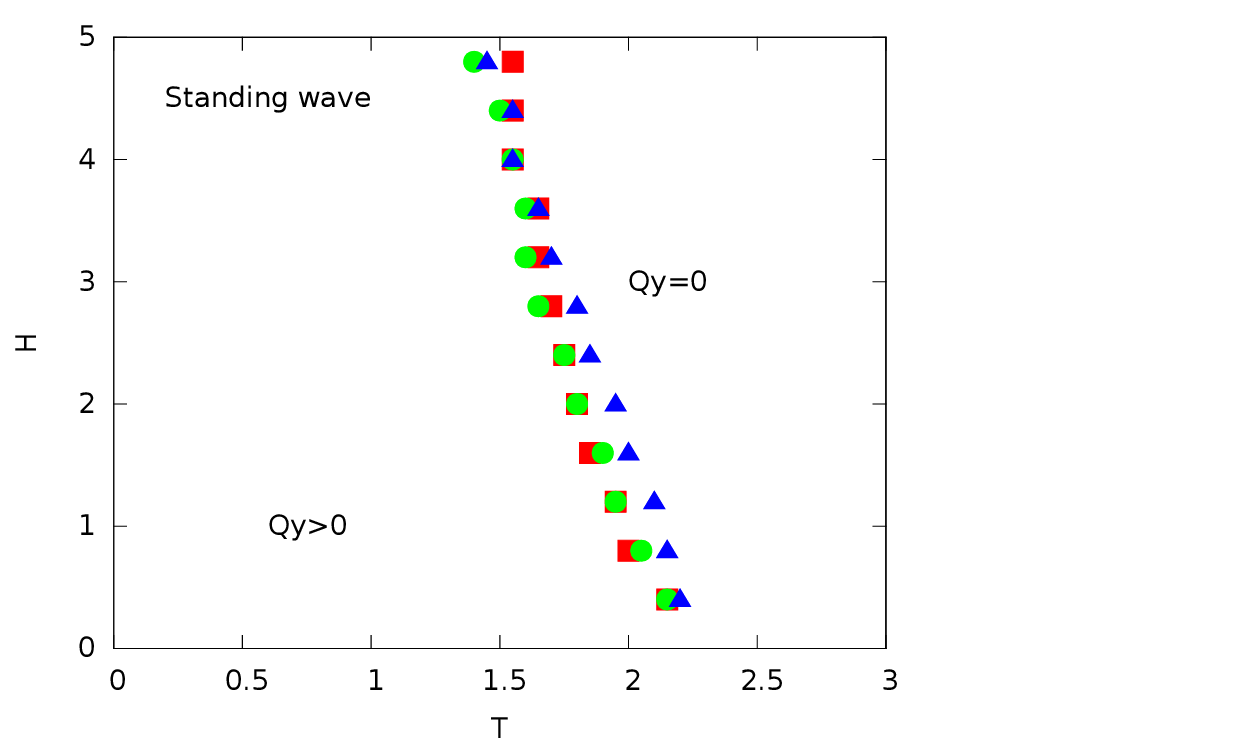}}
        
          \end{tabular}
\caption{The phase diagrams for {\it standing} wave. Different symbols correspond
to the different values of wavelength ($\lambda$). (Red square) $\lambda=20$, 
(Green bullet) $\lambda=10$ and (Blue triangle) $\lambda = 5$.}
\label{fig:standphase}
\end{center}
\end{figure}

Obtaining the values of dynamic transition temperatures from the
peak positions of $Var(Qy)$ and $C$ for different values of the amplitudes
($H$) the comprehensive phase diagrams are obtained and shown in 
Fig-\ref{fig:standphase}. 
Here also, the phase boundaries are drawn for three different
($\lambda=$ 20, 10 and 5) values of the wavelength of standing magnetic field
wave.
Unlike the case of propagating wave, the phase  
boundaries do not show any systematic variation with
 the wavelength of the standing magnetic
wave, in this simulational study.

\begin{figure}[h]
\begin{center}
\begin{tabular}{c}
        \resizebox{8cm}{!}{\includegraphics[angle=0]{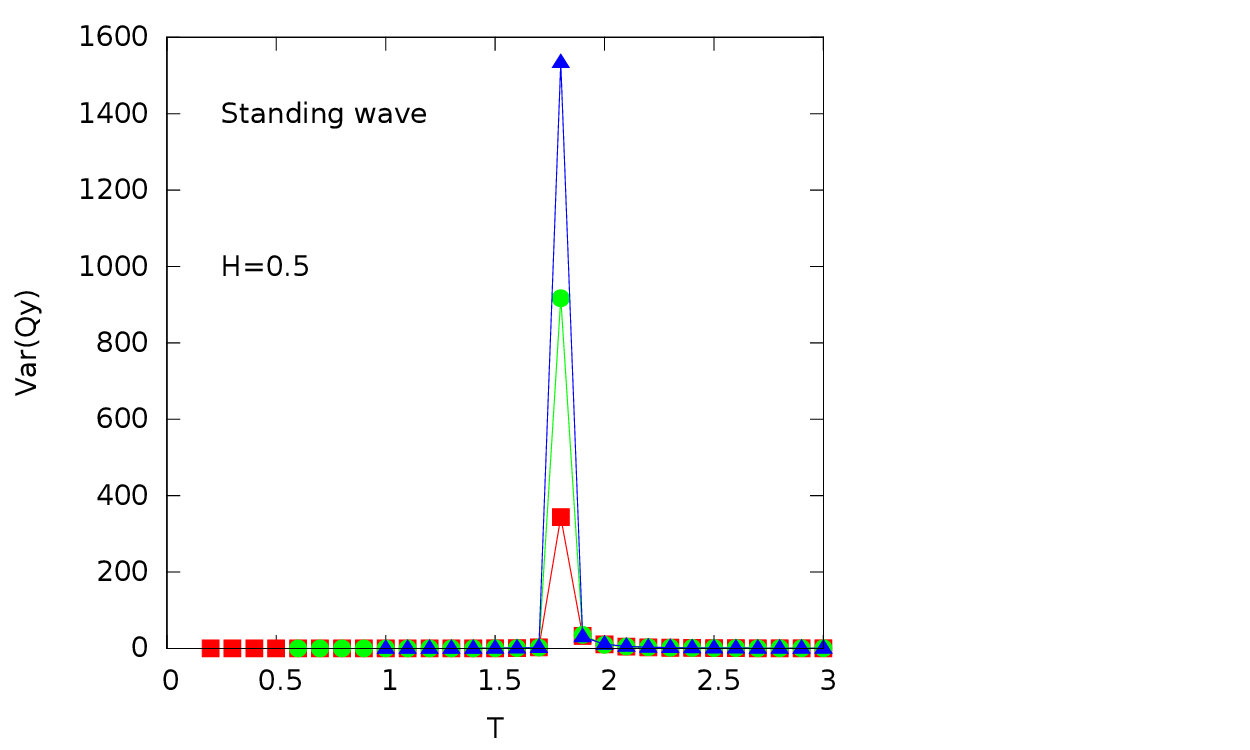}}
        
          \end{tabular}
\caption{The size dependence of the transition temperature in the 
case of {\it standing} wave. Different symbols correspond
to the different values of system size ($L$). (Red square) $L=20$, 
(Green bullet) $L=30$ and (Blue triangle) $L=40$. The wavelength 
of the standing wave $\lambda=10$ lattice unit, in all three cases.}

\label{fig:fss-sw}
\end{center}
\end{figure}

Here also, 
the transition temperatures are calculated from the peak position of
$Var(Qy)$ (studied as function of temperature) for three different values of
$L$ ($L=20,30$ and 40). Within the used accuracy ($\Delta T=0.02$ in this
simulation), it is observed that the positions of the peaks (for different $L$)
remain same (Fig-\ref{fig:fss-sw}).
In this case also, the peak height is observed to increase as $L$
increases showing the similar growth of critical correlation.

\vskip 1cm

\noindent {\bf IV. Summary}

The dynamical (or nonequilibrium) responses of classical 
XY ferromagnet driven by
linearly polarised propagating and standing magnetic field wave have been
investigated by Monte Carlo simulation using Metropolis algorithm in three
dimensions. The system is a simple cube of length (of each side)
 $L$ having periodic
boundary conditions in all three directions.

In the case of propagating magnetic field wave, with specified amplitude,
frequency and the wavelength, the system shows various dynamical responses
depending on the temperature. In the low temperature, the nonequilibrium
steady state is dynamically structured. The coherent motion of the spin
bands (or spin wave) was observed along the direction of the propagating 
magnetic field wave. In this phase, a nonzero net y component of magnetisation
was observed with vanishing average x component. 
The y component varies asymmetrically (about zero) with time.
The time averaged 
y component of the magnetisation over the full cycle of the propagating field
wave (i.e., the y component of dynamic order parameter) 
is nonzero. On the other hand, the high temperature, dynamical state 
is structureless or randomly oriented. As a result, both x and y components of
the time averaged magnetisations, over the full cycle 
of the propagating magnetic
wave, vanish. The x and y components of the
magnetisation, vary symmetrically (about zero) with time. A partially
(in y component only)
symmetry breaking dynamic nonequilibrium phase transition was observed
at any finite temperature. 
This symmetry breaking was observed independently from the distribution of
the angles of the spin vectors.
This transition temperature was found from the
temperature dependences of dynamic order parameter, its variance and the
dynamic specific heat. The dynamic transition was found to occur at lower
temperature with higher values of the amplitude of the propagating field.
A comprehensive nonequilibrium phase boundary was drawn in the plane formed
by the temperature and amplitude of the propagating field wave. The phase
boundary shrinks towards the low temperature region for longer wavelength of
the propagating magnetic field wave. A remarkable and interesting difference
in the nonequilibrium phases, with that observed in the discrete spin models
(Ising, Blume-Capel etc) \cite{ajay,blume}, is the absence of 
any pinned (or spin frozen) phase
in the low temperaure. The continuous spin models (like classical XY etc.)
does not require any threshold field to change the state of individual spin
which is essential for discrete spin systems. The phase boundaries were observed
to approach the equilibrium critical temperature (around 2.20$J/k$)
\cite{equcri} for vanishingly
small value of the field amplitude.

In the case of standing magnetic field wave, the low temperature nonequilibrium
phase is a standing spin wave. Here also the symmetry breaking nonequilibrium
phase transition was observed. The dynamic phase boundary was drawn. Unlike
the case of propagating magnetic wave, here the phase boundary does not 
show any significant dependence on the wavelength of the standing magnetic
field wave. However, like the case of propagating wave, the dynamic critical temperature
was observed to approach the equilibrium value (approximately 2.20$J/k$)
\cite{equcri}, as the value of the
field amplitude approaches zero.

Although the theoretical understanding of nonequilibrium phase transition is
not yet well developed, one may try to realise the effects as follows: The three
dimensional XY ferromagnet has an equilibrium critical temperature\cite{equcri} where the
time scale (relaxation time) diverges. Now consider, this system is being driven by
propagating magnetic field wave, which keeps the system away from the equilibrium.
This propagating magnetic field wave has a characteristic time (time period) and
a characteristic length (the wavelength). The nonequilibrium phase transition
is the outcome of the competition between the time scale of the driving field and the
intrinsic time scale (relaxation time) of the system. As the system is cooled from high
temperature (having random spin configuration)
the intrinsic time scale of the system increases. As a result, system gradually fails
to follow the driving field. Due to the relaxational delay, a phase lag, between the
response and the time dependent perturbation, developes. This gives rise to dynamical
ordering of the system. In this case, since the polarisation of the field is along
the X direction, the spin vector fails to follow the field. Consequently, the Y component
has a net average value, while the net X component becomes zero. And for higher values of
the field, the transition would occur at lower temperatures.

The phase boundary (particularly, in the case of propagating wave) was observed
to shrink (towards lower values of field strength and temperature). This may
be realised qualitatively as follows: The wavelength measures the
extension of the exposed
zone for the spin flips. As this area increases the lower value of temperature
and field strength would be adequate to set the dynamical order in the system.
However, this picture is incapable of explaining the complicated variation
of the phase boundary as function of the wavelength in the case of standing
wave. Further extensive investigations are required for clear understanding of this
behaviour of the phase boundary.

This study is an appeal to the experimentalists to see the effect of spin
dynamics in the ferromagnetic polycrystals 
like ${\rm Fe[Se_2CN(C_2H_5)_2]_2Cl}$ and ${\rm Zn[S_2CN(C_2H_5)_2]_2}$
which can be modelled by site diluted classical
XY system \cite{santos} with superexchange interactions. 
In the field of spintronics \cite{bader} it is also
important to know the response and thermodynamical behaviours of ferromagnetic
systems driven by intense optical perturbations.

\newpage

\newpage

\newpage

\newpage

\newpage

\newpage

\newpage

\newpage

\newpage

\newpage

\newpage

\newpage

\newpage

\newpage

\newpage
\end{document}